\documentclass{iopart}
\usepackage{amssymb}
\usepackage{graphicx}
\usepackage{geometry}
\usepackage{xcolor}
\usepackage{float}

\begin{document}
\title{Magnus Induced Diode Effect for Skyrmions in Channels with Periodic Potentials}
\author{J. C. Bellizotti Souza$^1$, 
            N. P. Vizarim$^{2,3}$, 
            C. J. O. Reichhardt$^4$, 
            C. Reichhardt$^4$
            and P. A. Venegas$^1$}
\address{$^1$ Departamento de F\'isica, Faculdade de Ci\^encias, Unesp-Universidade Estadual Paulista, CP 473, 17033-360 Bauru, SP, Brazil}
\address{$^2$ POSMAT - Programa de P\'os-Gradua\c{c}\~ao em Ci\^encia e Tecnologia de Materiais, Faculdade de Ci\^encias, Universidade Estadual Paulista - UNESP, Bauru, SP, CP 473, 17033-360, Brazil}
\address{$^3$ Department of Physics, University of Antwerp, Groenenborgerlaan 171, B-2020 Antwerp, Belgium}
\address{$^4$ Theoretical Division and Center for Nonlinear Studies, Los Alamos National Laboratory, Los Alamos, New Mexico 87545, USA
}
\ead{nicolas.vizarim@unesp.br}

\begin{abstract}
Using a particle based model, we investigate the skyrmion dynamical behavior in a channel where the upper wall contains divots of one depth and the lower wall contains divots of a different depth. Under an applied driving force, skyrmions in the channels move with a finite skyrmion Hall angle that deflects them toward the upper wall for $-x$ direction driving and the lower wall for $+x$ direction driving. When the upper divots have zero height, the skyrmions are deflected against the flat upper wall for $-x$ direction driving and the skyrmion velocity depends linearly on the drive. For $+x$ direction driving, the skyrmions are pushed against the lower divots and become trapped, giving reduced velocities and a nonlinear velocity-force response. When there are shallow divots on the upper wall and deep divots on the lower wall, skyrmions get trapped for both driving directions; however, due to the divot depth difference, skyrmions move more easily under $-x$ direction driving, and become strongly trapped for $+x$ direction driving. The preferred $-x$ direction motion produces what we call a Magnus diode effect since it vanishes in the limit of zero Magnus force, unlike the diode effects observed for asymmetric sawtooth potentials.  We show that the transport curves can exhibit a series of jumps or dips, negative differential conductivity, and reentrant pinning due to collective trapping events. We also discuss how our results relate to recent continuum modeling on a similar skyrmion diode system.
\end{abstract}

\maketitle

\vskip 2pc

\section{Introduction}

Diodes are crucial devices in modern electronics that can be used to control the electron flux. 
Electrical diodes permit easy current flow with small or zero resistance
in one direction, and hinder the flow
with a high or infinite resistance in the other direction
\cite{Kitai11}. This unidirectional diode flow
has inspired investigations in several other branches of physics, including
photonics
\cite{tocci_thinfilm_1995,scalora_photonic_1994,wang_optical_2013,sciamanna_physics_2015},
heat-transfer dynamics \cite{li_thermal_2004,martinez-perez_rectification_2015}, fluidics \cite{mates_fluid_2014,shou_all_2018}, and magnetism
\cite{zhao_ferromagnetic_2020,wang_magnetic_2020,tulapurkar_spin-torque_2005,song_spin-wave_2021,Jung21,Feng22,fang_giant_2016,harrington_practical_2009,lyu_superconducting_2021}.

The basic unidirectional flow property of diodes can be implemented in several different ways depending on the physical system.
For example, in  
electrical diodes, electrons may flow in one direction and not the other depending on the density of
positive or negative charge carriers present in a semiconducting material. Analogously, for thermal diodes the 
heat has preferential directions of flow.
For magnetic diodes that affect the flow of 
superconducting vortices or magnetic textures, the diode mechanism can be controlled by an external 
magnetic field, a driving current, or by using Hall effects.
In the case of superconducting vortices, a diode effect can be 
produced with a combination of asymmetric pinning potentials and oscillating drives that cause the vortices to undergo ratcheting motion
\cite{lu_reversible_2007,wordenweber_guidance_2004,van_de_vondel_vortex-rectification_2005}. 
It was also shown that dc drives can produce diode effects
since multiple vortices under the 
influence of constriction arrays can flow easily in one direction and clog in the other 
\cite{olson_reichhardt_vortex_2013,reichhardt_jamming_2010}.
More recently, increasing effort has focused on developing
ways to control
the motion of magnetic textures such as magnetic skyrmions,
since their high stability and 
reduced size makes them ideal for use in novel technological devices.

Skyrmions are a topologically stable particle-like object composed of spins
pointing
in all directions
wrapping a sphere
\cite{muhlbauer_skyrmion_2009,fert_skyrmions_2013,fert_magnetic_2017}.
They can be set in motion by the application of a spin polarized current that
is larger than depinning
threshold, similar to superconducting vortices 
\cite{jonietz_spin_2010,schulz_emergent_2012,yu_skyrmion_2012,olson_reichhardt_comparing_2014}.
The main difference between skyrmions and other overdamped particles is that skyrmions exhibit strong 
gyroscopic effects,
so there is a Magnus term in their equation of motion that results in the 
appearance of a
skyrmion Hall effect \cite{nagaosa_topological_2013,iwasaki_universal_2013,fert_magnetic_2017}.
The 
Magnus term produces a force that is perpendicular to the skyrmion velocity,
and it has been suggested that this
is the main cause of the low depinning thresholds observed for skyrmions.
When skyrmions flow in a pure sample without defects,
they move at an intrinsic 
angle, known
as the intrinsic skyrmion Hall angle $\theta_{sk}^{int}$, to the
applied drive direction
\cite{nagaosa_topological_2013,fert_magnetic_2017,jiang_direct_2017}.
The magnitude of this angle
depends on the ratio of the Magnus term to the damping 
term.
Experimentally, skyrmion Hall angles have been observed that span the range
from a few degrees up to 
very large angles, depending on the system parameters and
the size of the skyrmions 
\cite{jiang_direct_2017,zeissler_diameter-independent_2020,litzius_skyrmion_2017}.

The theoretical idea
of a skyrmion diode was recently proposed by Jung {\it et al.}
\cite{Jung21}, who
used continuum-based simulations to demonstrate
a skyrmion diode device exhibiting
unidirectional skyrmion transport.
The device is designed as a 
nanostrip with asymmetric edges, where there is a pinning site in one of the edges.
The skyrmions 
experience different pinning potentials depending on the direction of the applied transport force, resulting in a
diode effect.
In another theoretical proposal for a skyrmion diode
device, Feng {\it et al.} \cite{Feng22} consider
a rectangular channel
connected to a smaller strip. For driving applied in one direction,
the skyrmion is trapped by the intrinsic skyrmion Hall 
angle in the smaller region, where it annihilates.
If the drive is applied in the other direction,
the skyrmion flows in a straight line 
following the edge of the sample without being annihilated. This creates a clear diode effect where 
there is absolutely no flow in one direction and ordinary flow in the other direction.
These 
papers only treated the case of single skyrmion dynamics.
Additional work on similar geometries
for systems with modified edges
describes
the creation of NAND and NOR gates \cite{Shu22}.
There is also an interesting proposal to create a
skyrmion diode using spin waves in the presence of a transverse magnetic 
field \cite{song_spin-wave_2021}.

In this work we investigate the collective behavior of skyrmions under the influence of asymmetric 
potentials and dc drives. We conduct simulations of two types of channel geometries. The first geometry, labeled S1, is similar 
to that used by Feng {\it et al.} \cite{Feng22}, where the
top wall of the channel is featureless
and the bottom wall contains divots that can trap skyrmions.
The second geometry, labeled S2, has divots on both 
the upper and lower walls, but the divots on the upper wall are
shallower than those on the bottom.
Our results show that due to the asymmetry in the potentials, 
the skyrmions exhibit a preferred direction of motion, leading to a diode effect.
In the high asymmetry S1 sample, for 
drives applied along the $+x$ direction,
the intrinsic Hall angle causes the skyrmions to become trapped in the 
lower divots. On the other hand, when the drive is applied along
the $-x$ direction, the intrinsic Hall angle pushes the skyrmions
against the featureless wall, where they can travel with high velocities.
For the lower asymmetry S2 sample,
when the drive is applied along the $+x$ direction,
the skyrmions become trapped in the deeper lower divots, and
their motion is hindered.
In contrast, for drives applied along the $-x$ direction, the 
skyrmions are brought into contact with the upper shallower divots, which
do not impede the motion as much.
Our work not only complements the results of the continuum 
based simulations, but also allows us to consider large assemblies
of skyrmions in order to study collective effects
and velocity-force curves more clearly.
For example, we find that the velocity-force curves
show not only diode effects
but also negative differential conductivity where the
magnitude of the velocity 
drops with increasing drive, as well as reentrant pinning
where the system is strongly 
pinned at high drives but flows easily at low drives.  

We emphasize that the geometry we consider is distinct from the
asymmetric potentials such as sawtooth shapes used to
create ratchet effects. In those systems,
the potential barrier is higher in one direction than the other,
so that under an ac drive there is dc flow only in the easy direction.
Ratchet effects for skyrmions in
such asymmetric potentials have been studied previously \cite{reichhardt_magnus-induced_2015,ma_reversible_2017,souza_skyrmion_2021}.
In these systems, ratchet and diode effects occur even in the absence of Magnus forces, similar to what is found for vortex ratchets
\cite{lu_reversible_2007,wordenweber_guidance_2004,van_de_vondel_vortex-rectification_2005}. When a Magnus force is
present, additional symmetry breaking arises
due to the chirality of the Magnus term, making it possible to create
ratchet effects even in a spatially symmetric potential
\cite{gobel_skyrmion_2021}. For the geometries we consider,
diode and ratchet effects do not occur in the overdamped limit.
Beyond skyrmions, our results should be relevant for creating
diode devices for any type of particle system
with a Hall deflection,
such as chiral active matter with odd viscosity \cite{Banerjee17} or
charges in a magnetic field
including Wigner crystals \cite{Reichhardt21} and dusty plasmas \cite{Melzer21}.

\section{Simulation}

\begin{figure}[h]
\begin{center}
  \includegraphics[width=0.8\columnwidth]{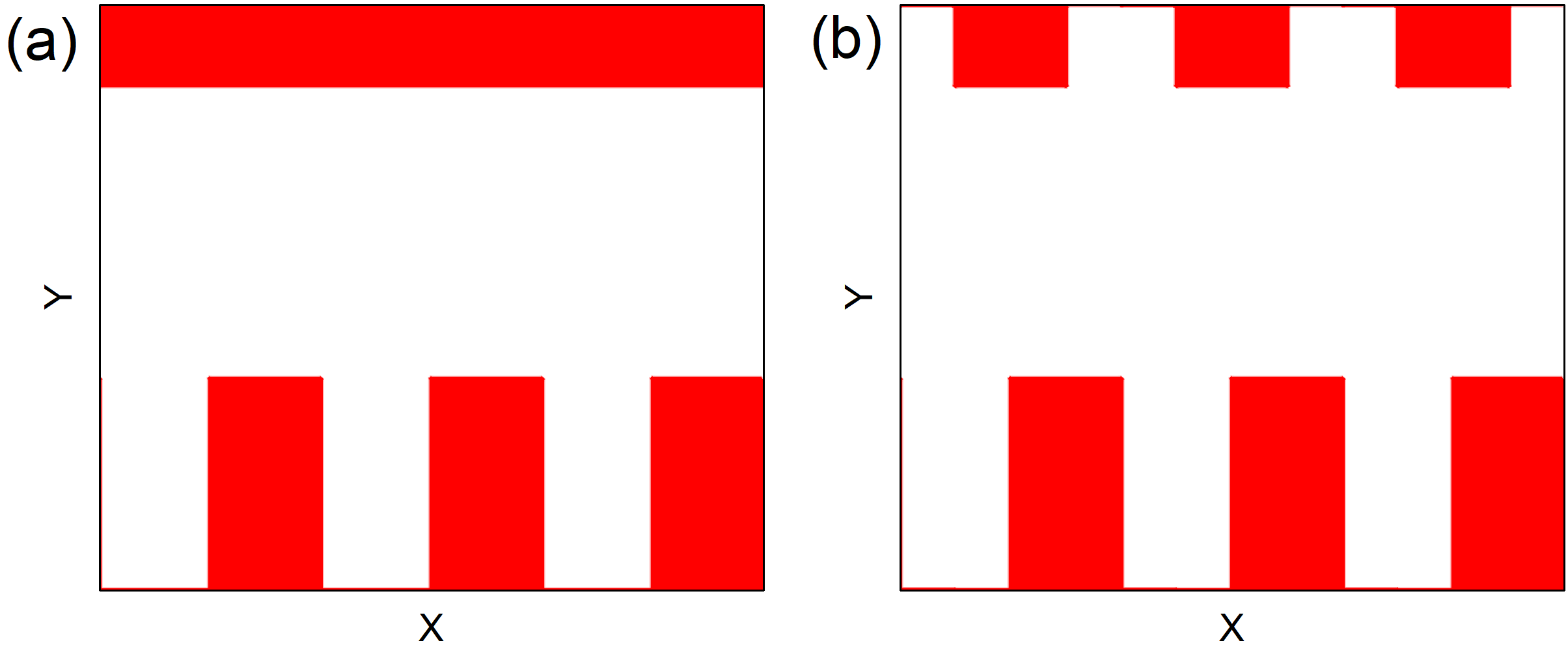}
  \end{center}
\caption{Images of the samples used in this work. Red areas are
forbidden to the skyrmions.
(a) The high asymmetry  geometry S1.
(b) The lower asymmetry geometry S2, where
the lower divots have a depth of $13\xi$ and the upper divots
have a depth of $5\xi$.
    }
    \label{fig1}
\end{figure}

We simulate the dynamics of multiple interacting skyrmions
in the presence of asymmetric channel potentials S1 and S2,
illustrated in Fig. \ref{fig1}.
The system is of size
$L_x\times L_y$, with $L_x=L_y=36\xi$, and
there are periodic boundary conditions
only in the $x$ direction.
Here $\xi$ is the size of the skyrmion core.
The samples contain prohibited areas for the skyrmions,
shown in red in Fig.~\ref{fig1},
where the energy potential is
very high.
As a result, skyrmions can only exist stably in the white regions
where the sample is clean.
The separation between the red and white regions is modeled with
repulsive barrier walls
that confine the skyrmions to the white region.
In addition to interacting with the barrier walls,
the skyrmions interact
with each other and also with an applied dc current.
We initially consider the simpler geometry labeled S1, shown in
Fig.~\ref{fig1}(a),
where the upper channel wall is flat and the lower wall contains divots,
giving a high asymmetry. 
We analyze the skyrmion dynamical behavior under
dc transport forces applied along the positive and negative $x$ direction.
We next consider the lower asymmetry geometry S2, shown in
Fig.~\ref{fig1}(b),
where both the upper and lower walls contain divots, but the divots on the
upper wall are shallower than those on the lower wall.
The skyrmion density is $n=N_{sk}/A_w$,
where $A_w = L_x L_y - A_r$
is the area of the white region,
$A_r$ is the area of the prohibited red region, and
$N_{sk}$ is the number of skyrmions.
The lower divots
in both S1 and S2 have dimensions of
$13\xi \times 6\xi$,
and the upper divots in S2 are of size $5\xi \times 6\xi$.
Unless otherwise noted,
we consider samples with
$N_{sk}=26$ skyrmions, so that $n_{S1}=0.0295/\xi^2$ and 
$n_{S2}=0.0268/\xi^2$. 

To simulate the skyrmion behavior we use 
a particle-based model for skyrmions \cite{Lin13}, given by:

\begin{equation}\label{eq}
    \alpha_d\mathbf{v}_i+\alpha_m\hat{z}\times\mathbf{v}_i=
    \mathbf{F}^{SS}_i+\mathbf{F}^{W}+\mathbf{F}^{D} \ .
\end{equation}

Here, $\mathbf{v}_i$ is the velocity of $i$th skyrmion. The first term on the left
side with prefactor $\alpha_d$ is the damping term, which arises from spin precession and dissipation of electrons
in the skyrmion core. The second term
on the left with prefactor $\alpha_m$ is the Magnus force
from gyroscopic effects. The Magnus force is oriented 
perpendicular to
the skyrmion velocity. Throughout this work
we use the normalization $\alpha_d^2+\alpha_m^2=1$.
The first term on the
right side is the skyrmion-skyrmion repulsive interaction force, given by
$\mathbf{F}_i^{ss}= \sum_{i}^{N_{sk}} K_{1} 
(r_{ij}/\xi) {\mathbf{\hat{r}}_{ij}}$, where $K_1$ is the modified 
Bessel function of the second kind,
$r_{ij}=|\mathbf{r}_i - \mathbf{r}_j|$ is the distance between skyrmions $i$ and 
$j$, and ${\mathbf{\hat{r}}_{ij}}=(\mathbf{r}_i - \mathbf{r}_j)/r_{ij}$.
For $r_{ij}>6.0$ the skyrmion-skyrmion interaction is cut
off for computational efficiency.
The second term on the right side is the interaction between skyrmions and the repulsive walls, given by
$\mathbf{F}^{W}=2r_{iw}U_0/a_0^2\exp\left(-r_{iw}^2/a_0^2\right)\hat{\mathbf{r}}_{iw}$, where $U_0=30.0$ is 
the 
potential strength, $a_0=0.05\xi$ is the wall thickness, 
and $r_{iw}$ is the distance between skyrmion $i$ and wall $w$. 
The values of $U_0$ and 
$a_0$ 
were chosen large enough to prevent the skyrmions from
crossing the barrier walls, so that the skyrmions
remain trapped in the 
white regions shown in Fig. \ref{fig1}. 
The last term on the right in equation \ref{eq}
is the force from the external drive, 
$\mathbf{F}^{D}=F^D\hat{\mathbf{d}}$, 
where $\hat{\mathbf{d}}=\pm{\bf \hat x}$
is the applied current direction.
We increase $F^{D}$ in small steps of $\delta F^{D}= 0.003$
and spend $5\times 10^{5}$
simulation time steps at each drive increment.
We normalize all distances by
the screening length $\xi$ and the
average velocities are obtained with
$\left\langle V_x\right\rangle=\left\langle \mathbf{v} \cdot \widehat{\rm 
{\bf{x}}}\right\rangle$ and $\left\langle V_y\right\rangle = \left\langle \mathbf{v} \cdot
\widehat{\rm {\bf{y}}}\right\rangle$,
where the averages are taken over the
simulation time steps spent at each drive increment.

\section{System S1 with higher asymmetry}

We first analyze system S1 where there is a high asymmetry 
between the upper and lower part of the sample,
as shown in Fig.~\ref{fig1}(a). 
To measure the dynamic behavior of skyrmions
inside this geometry, we add $N_{sk}=26$ skyrmions to the
sample and perform simulated annealing
until a ground state is reached.
We then apply an external driving force along
the $-x$ (easy) and $+x$ (hard) directions.
Since skyrmions exhibit finite Hall angles, 
it is expected that skyrmions might be pushed towards
the smooth upper wall or toward the divots depending on
the direction of the applied drive.
Initially we explore
the simpler case of a drive applied along the 
easy direction.

\subsection{Easy direction ($-x$) driving}

\begin{figure}[h]
\centering
\includegraphics[width=0.6\columnwidth]{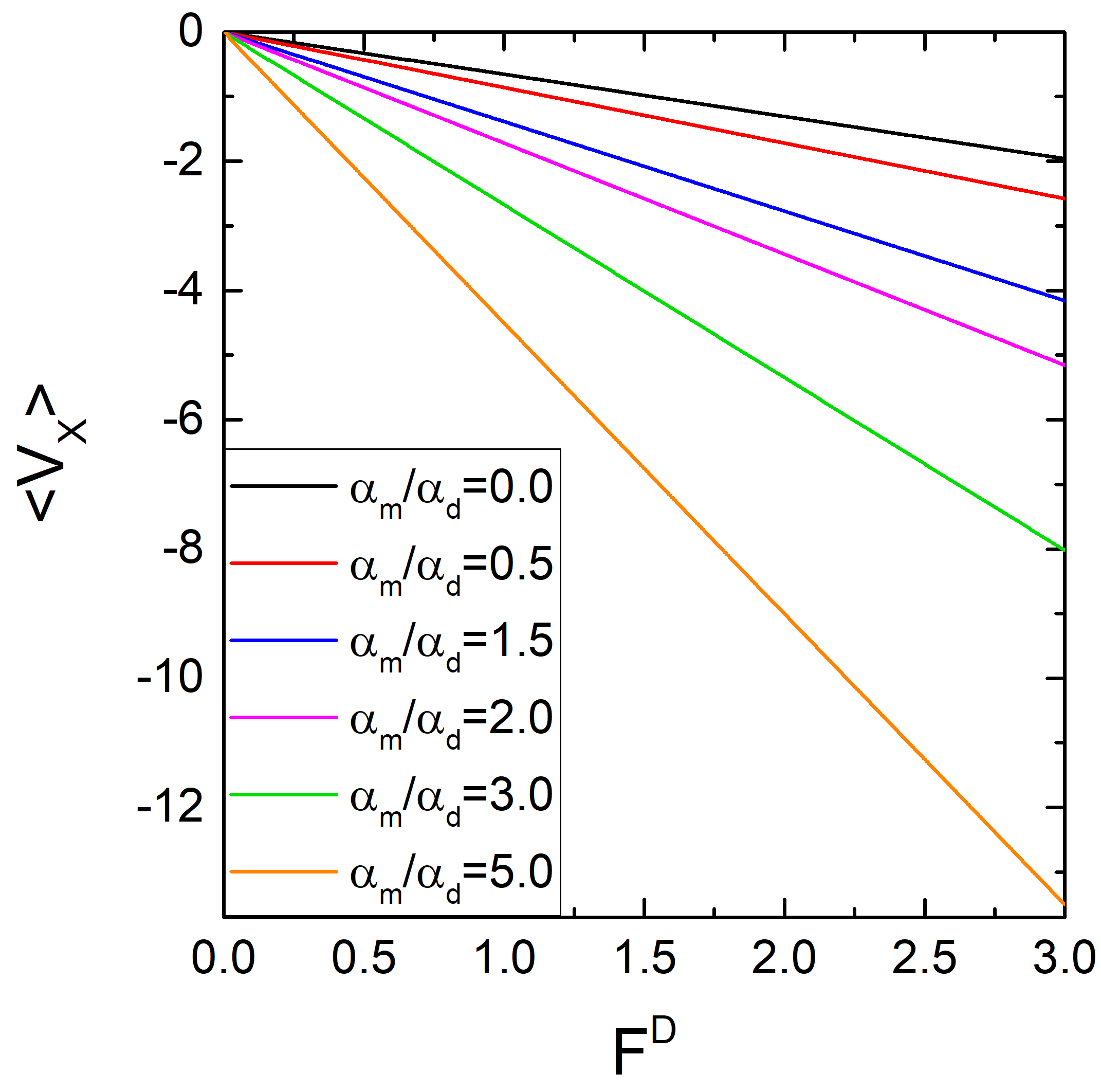}
\caption{$\left\langle V_x\right\rangle$ vs $F^D$ for selected values of $\alpha_m/\alpha_d$
for the high asymmetry system S1
under an easy direction applied drive
with ${\bf \hat d}=-{\bf \hat x}$.
In all cases, the
magnitude of the velocity increases linearly with the applied drive.}
    \label{fig2}
\end{figure}

\begin{figure}[h]
\centering
\includegraphics[width=0.45\columnwidth]{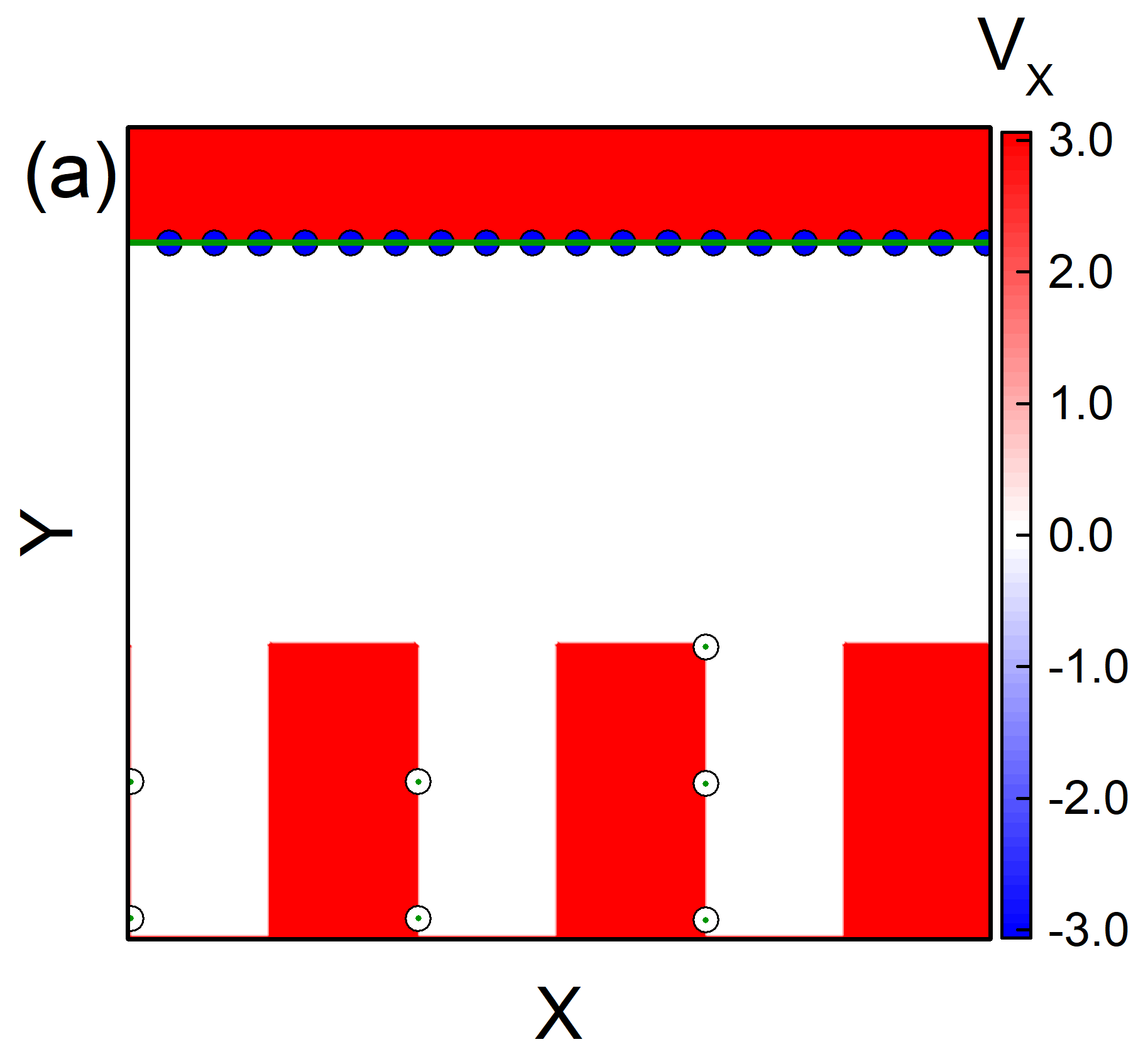}
\includegraphics[width=0.45\columnwidth]{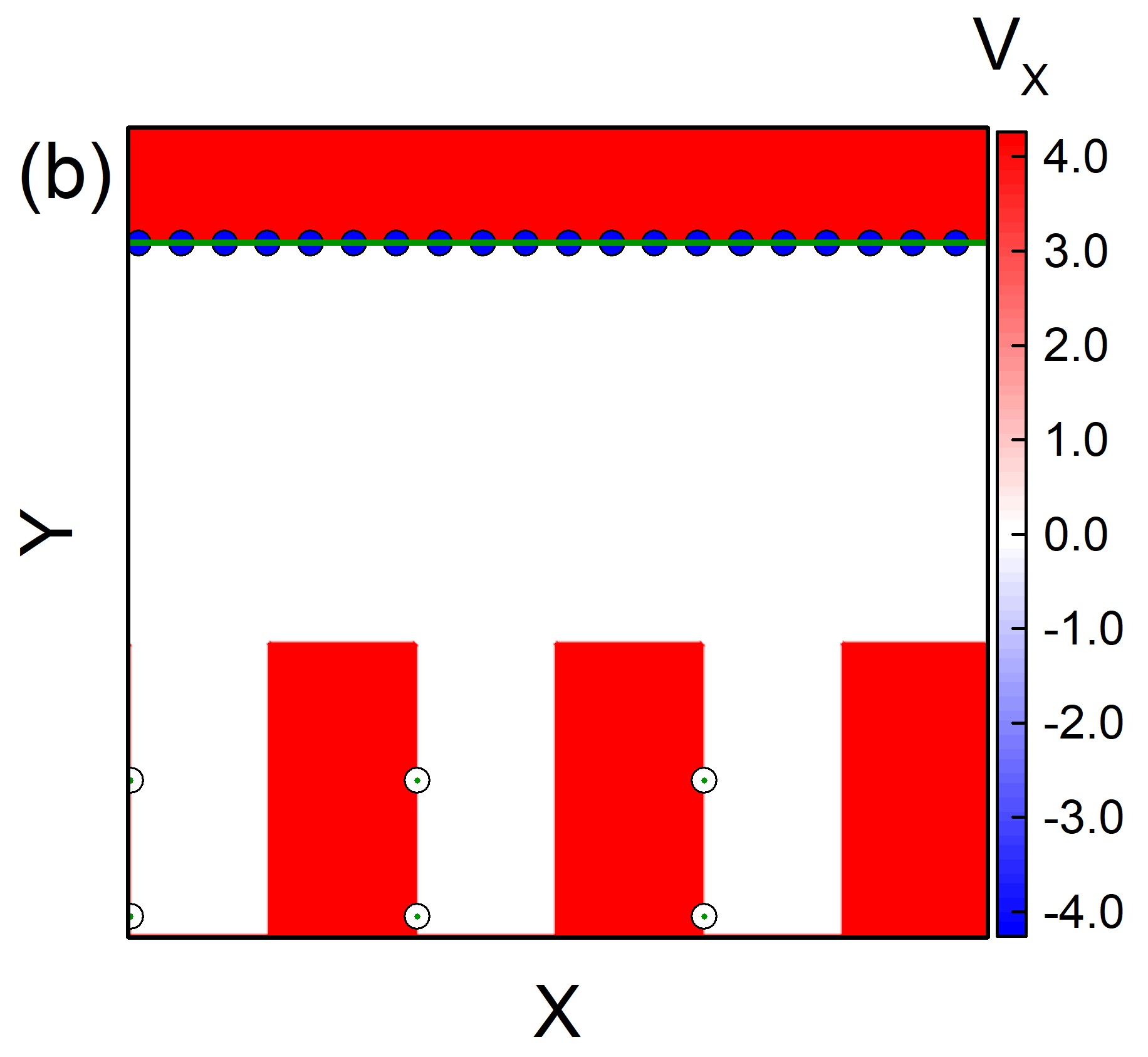}
\caption{Skyrmion starting positions (circles)
and trajectories (green lines)
for ${\bf \hat d}=-{\bf \hat x}$ or easy direction driving
in the high asymmetry system S1 with
$F^D=3.0$
at (a) $\alpha_m/\alpha_d=0.2$ and (b) $\alpha_m/\alpha_d=1.0$.
Red areas are forbidden to the skyrmions.
Each skyrmion instantaneous velocity, $V_x$, 
is represented by a color scale attached to the plot, where
white are skyrmions with null velocity, blue negative and red positive velocities. 
}
    \label{fig3}
\end{figure}

In Fig.~\ref{fig2} we
plot $\left\langle V_x\right\rangle$ as a function of $F^D$ for the
system shown in Fig.~\ref{fig1}(a)
for selected 
values of $\alpha_m/\alpha_d$ under a drive applied along
${\bf \hat d}=-{\bf \hat x}$.
The magnitude of the velocity increases linearly with increasing
drive since,
in this configuration, skyrmions are pushed to the
upper part of the sample and encounter the
straight repulsive barrier wall. The skyrmions can not
surpass this barrier,
so they slide against the wall in the $-x$ direction.
Note that the speed of
the skyrmion motion changes as $\alpha_m/\alpha_d$
varies.
In the hypothetical case of a skyrmion with no
Magnus force, $\alpha_m/\alpha_d=0$,
similar behavior appears but the magnitude of the
velocity is diminished.
The mechanism of the skyrmion sliding along
repulsive walls due to the Magnus force
was explained recently in Ref.~\cite{souza_clogging_2022}.
In Fig.~\ref{fig3} we plot
selected skyrmion trajectories for the system in
Fig.~\ref{fig2} at $F^D=0.3$.
A sample with $\alpha_m/\alpha_d=0.2$,
where the intrinsic Hall angle $\theta_{sk}^{\rm int}=11.31^{\circ}$,
appears in Fig.~\ref{fig3}(a), while a system with
$\alpha_m/\alpha_d=1.0$  and $\theta_{sk}^{\rm int}=45^{\circ}$ is
shown in Fig.~\ref{fig3}(b).
As expected, a portion of the skyrmions remain
trapped in the lower divots, while the remaining skyrmions
flow along the upper straight wall due to the skyrmion Hall effect.

\subsection{Hard direction ($+x$) driving}

\begin{figure}[h]
\centering
\includegraphics[width=0.32\columnwidth]{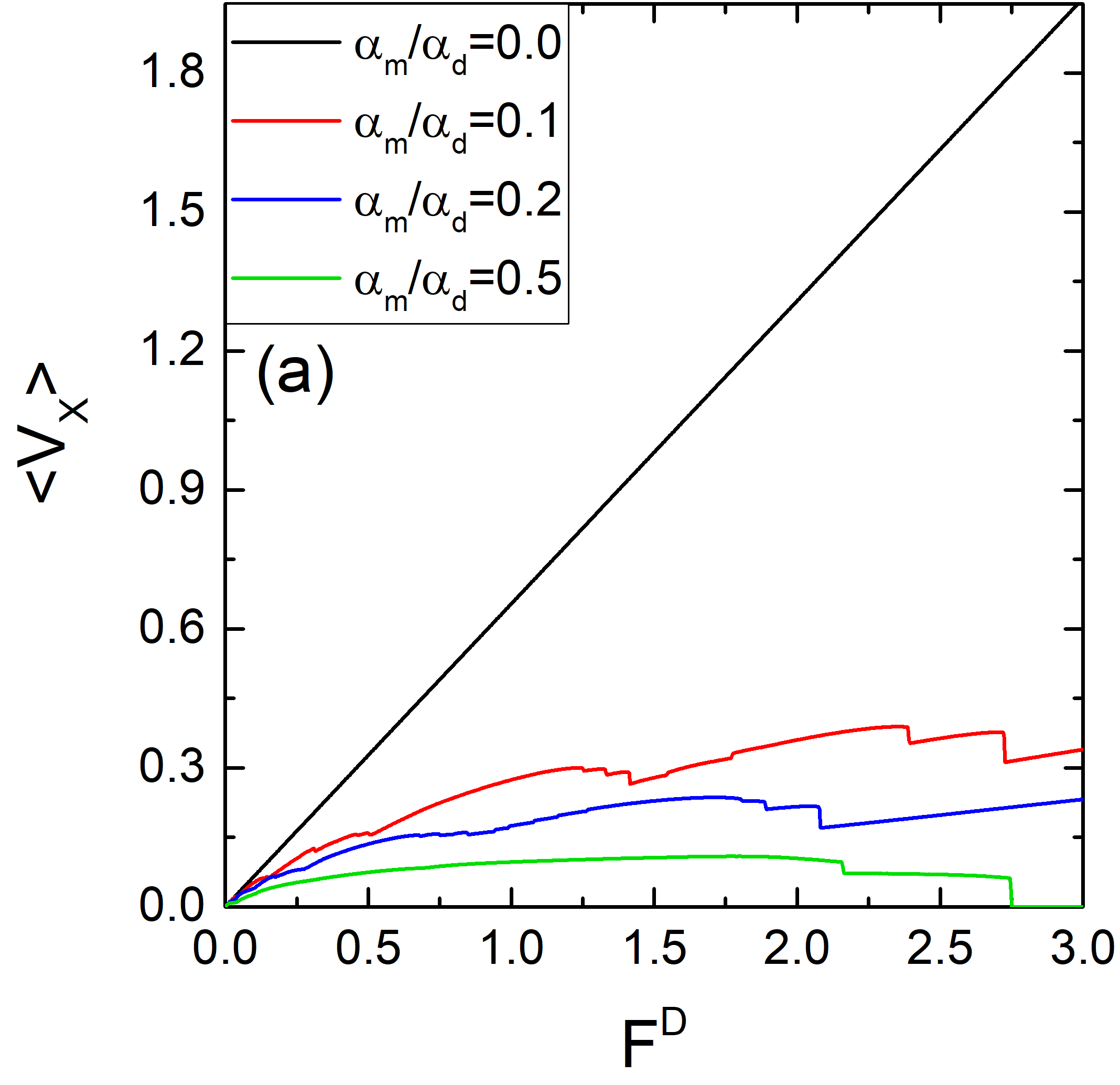}
\includegraphics[width=0.32\columnwidth]{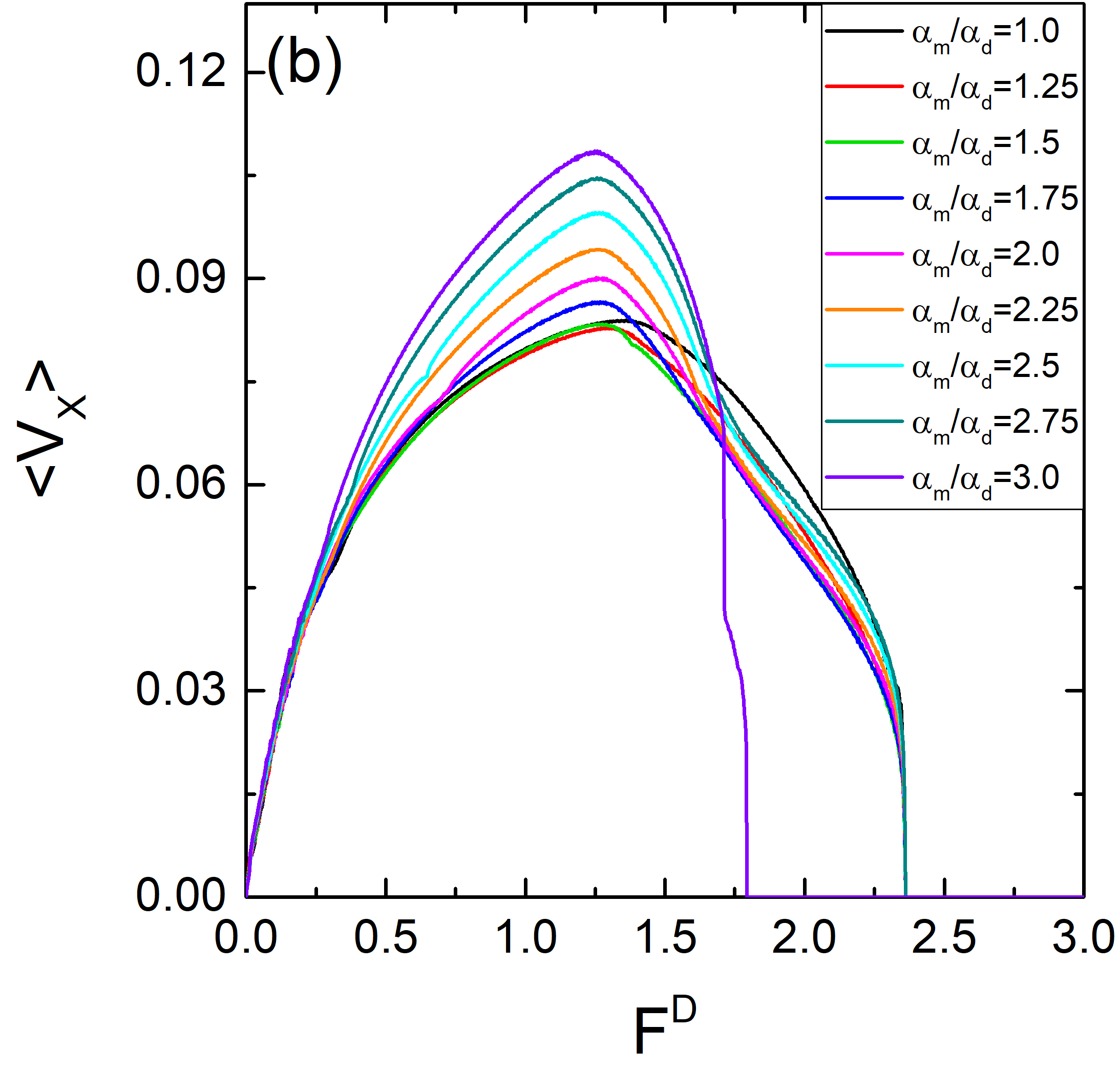}
\includegraphics[width=0.32\columnwidth]{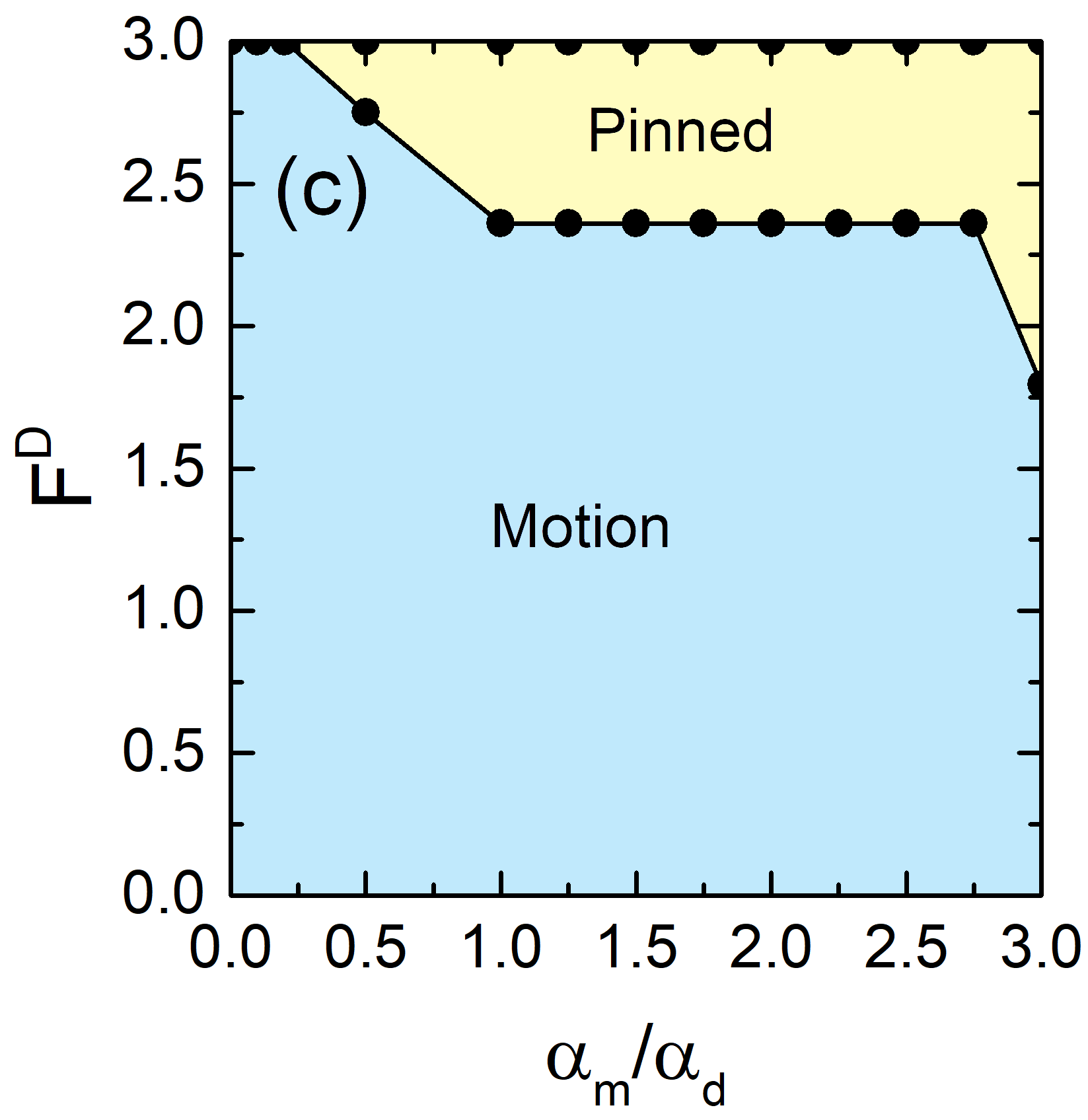}
\caption{
Results for hard direction driving along ${\bf \hat d}=+{\bf \hat x}$
for the high asymmetry system S1.
(a) $\left\langle V_x\right\rangle$ vs $F^D$ for
several values of $\alpha_m/\alpha_d$ in the
weak Magnus force regime,  $0 \leq \alpha_m/\alpha_d \leq 0.5$. 
(b)
The same for the intermediate Magnus force
regime, $1.0 \leq \alpha_m/\alpha_d \leq 3.0$.
(c) Dynamical phase diagram as a function of $F^D$ vs
$\alpha_m/\alpha_d$ showing the reentrant pinned phase (yellow) and the
moving phase (blue).}
\label{fig4}
\end{figure}

We next consider applying the drive along the hard
direction, ${\bf \hat d}=+{\bf \hat x}$.
In this case, the
skyrmion Hall angle pushes the skyrmions towards the divots.
In Fig. \ref{fig4}(a,b) we plot $\left\langle V_x\right\rangle$ as a function of $F^D$ for the
system shown in Fig.~\ref{fig1}(a) at several 
values of $\alpha_m/\alpha_d$.
The results are separated into two plots for better visualization of the
data. Figure~\ref{fig4}(a) shows weak Magnus forces,  $0 \leq \alpha_m/\alpha_d \leq 0.5$, while Fig.~\ref{fig4}(b) shows
intermediate Magnus forces, $1.0 \leq \alpha_m/\alpha_d \leq 3.0$.
In Fig. \ref{fig4}(c) we construct a dynamic phase diagram as a function of
$F^D$ versus $\alpha_m/\alpha_d$.

\begin{figure}[h]
\centering
\includegraphics[width=0.32\columnwidth]{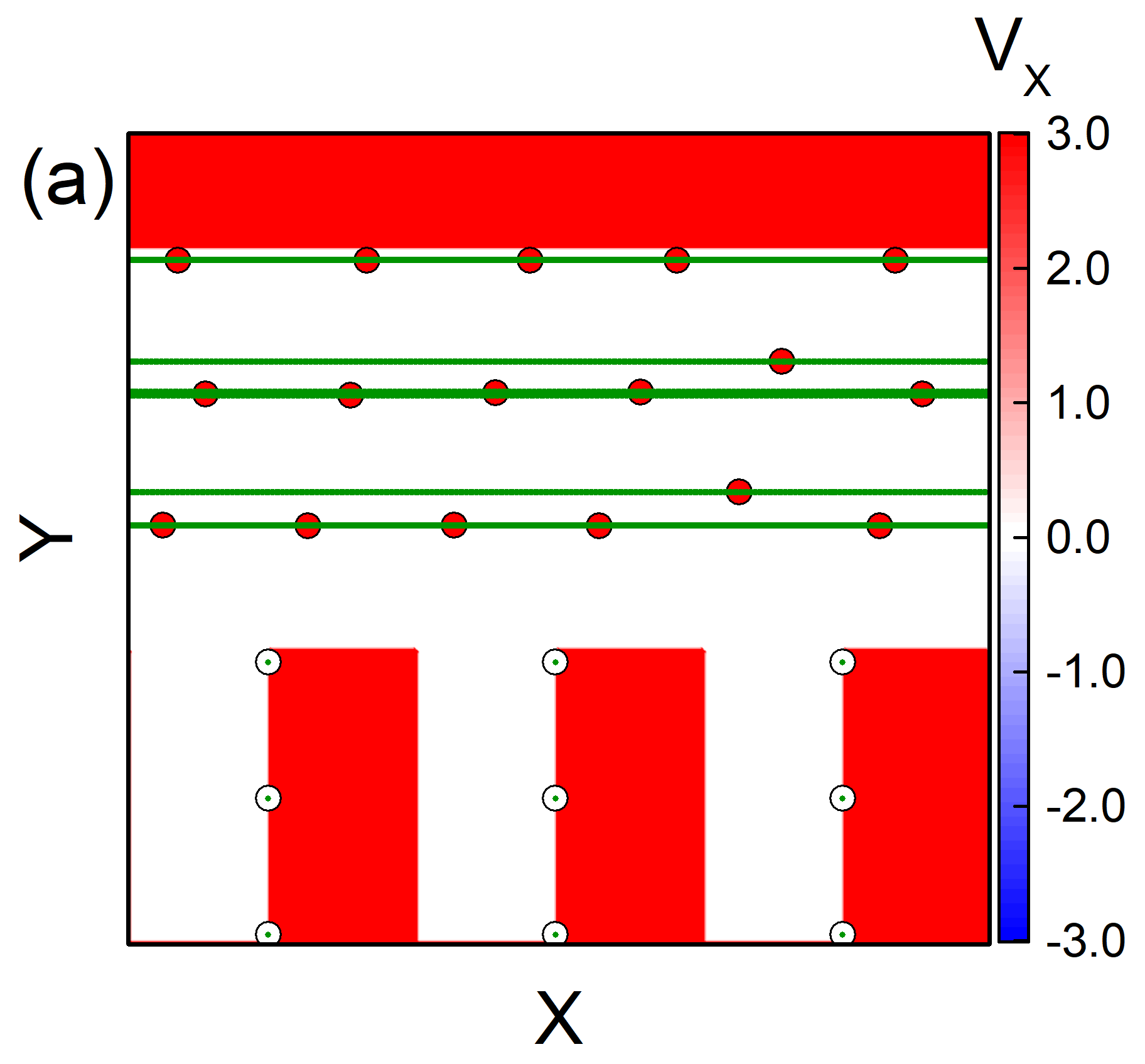}
\includegraphics[width=0.32\columnwidth]{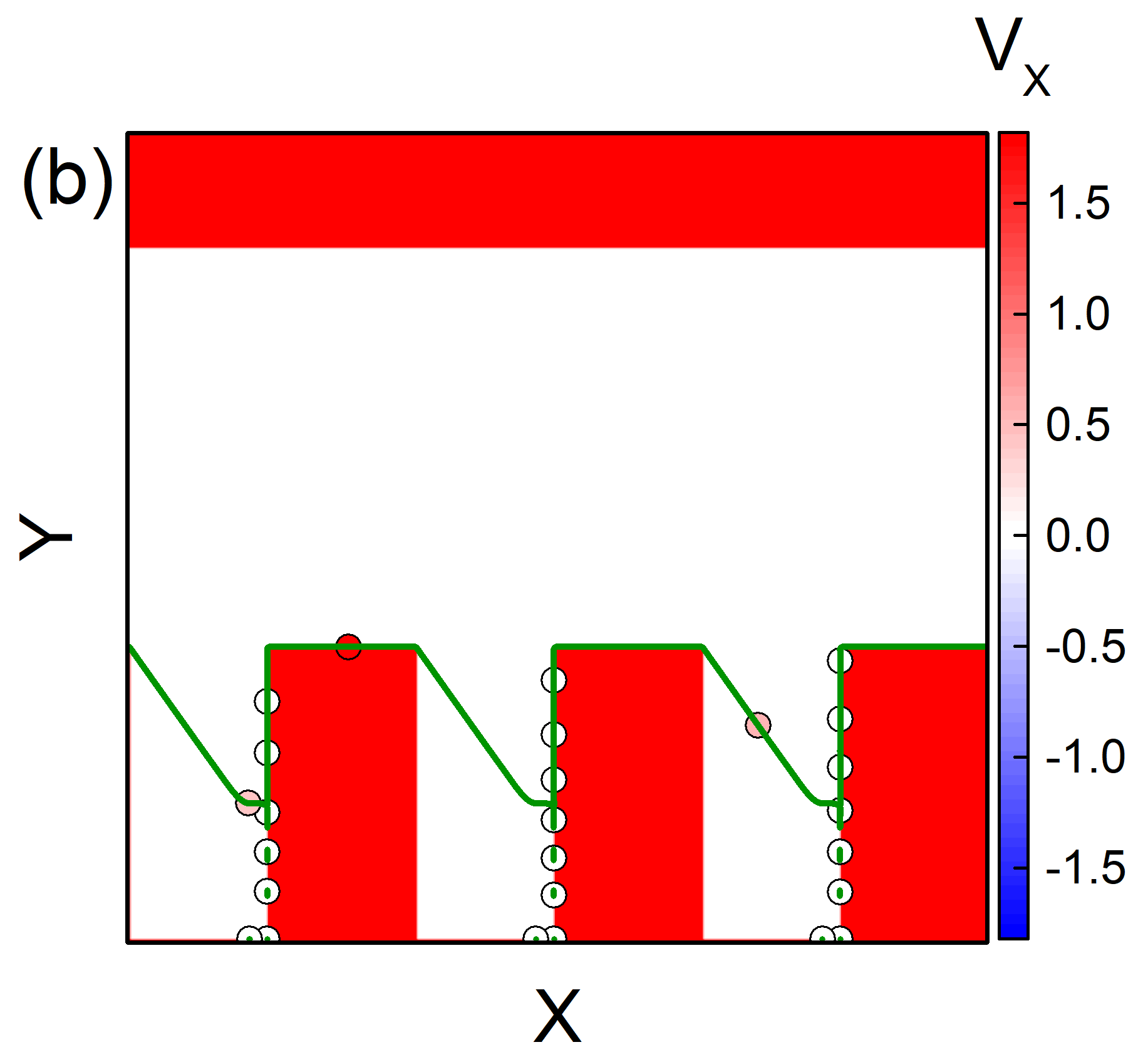}
\includegraphics[width=0.32\columnwidth]{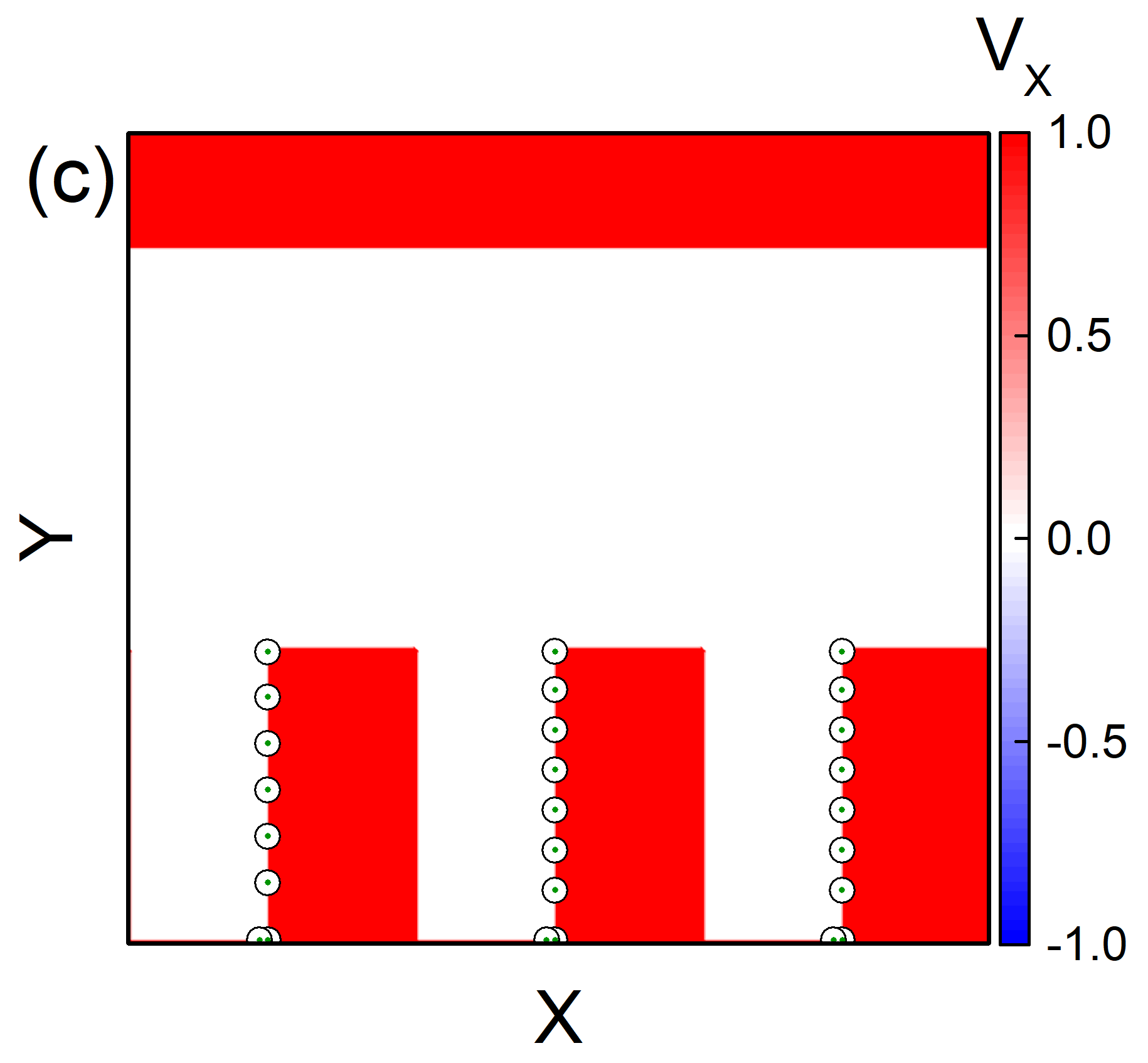}
\caption{Skyrmion positions (black dots) and trajectories
(green lines) for ${\bf \hat d}=+{\bf \hat x}$ or hard direction
driving in the high asymmetry S1 system shown in Fig.~\ref{fig1}(a). 
Red areas are forbidden to the skyrmions.
Each skyrmion instantaneous velocity, $V_x$, 
is represented by a color scale attached to the plot, where
white are skyrmions with null velocity, blue negative and red positive velocities. 
(a) $\alpha_m/\alpha_d=0$ and $F^D=3.0$,
where skyrmions do not exhibit the skyrmion
Hall effect due to the absence of a Magnus term,
but instead flow along the direction of the applied drive.
(b) $\alpha_m/\alpha_d=1.5$ and $F^D=1.0$,
where a portion of the skyrmions in the lower part of the sample flow
while the other skyrmions remain trapped.
(c) $\alpha_m/\alpha_d=1.5$ and $F^D=2.5$,
where there is a pinned phase.
}
    \label{fig5}
\end{figure}

For the low Magnus force regime in Fig.~\ref{fig4}(a),
we find when the Magnus force is completely absent,
$\alpha_m/\alpha_d=0$, the presence of divots 
has no influence on the dynamic behavior of the skyrmions.
As illustrated in Fig.~\ref{fig5}(a), the skyrmions
travel in a straight line
with $\theta_{sk}^{\rm int}=0^{\circ}$,
and 
their velocity increases linearly with increasing applied drive.
That is, without a Magnus force,
this system exhibits no easy or hard direction of motion.
On the other hand, when the Magnus force is present, the 
skyrmion Hall effect causes the
skyrmions to move towards the divots, resulting in a
significant reduction 
of the average skyrmion velocity $\langle V_x \rangle$ as
shown in Fig.~\ref{fig4}(a).
Thus the possibility to control the skyrmion velocity
using the type of geometry proposed in this work is 
only effective for particles where $\alpha_m/\alpha_d>0$,
such as ferromagnetic skyrmions.
For intermediate values of Magnus forces,
shown in Fig.~\ref{fig4}(b), the skyrmion motion ceases
above $F^D>2.36$ for most values of $\alpha_m/\alpha_d$.
Note that for $\alpha_m/\alpha_d=0.5$ in
Fig.~\ref{fig4}(a),
there is also a region where $\langle V_x \rangle=0$ for $F^D>2.75$. 
In Fig.~\ref{fig5}(b) we show a snapshot of the skyrmion
motion
for the system with $\alpha_m/\alpha_d=1.5$ or $\theta_{sk}^{\rm int}=56.31^\circ$
at $F^D=1.0$, and
in Fig.~\ref{fig5}(c) we show the same system at $F^D=2.5$.
When $F^D=1.0$, the skyrmions flow through the sample in an
ordered fashion.
Skyrmions accumulate on the left side of each divot and,
due to a combination of skyrmion-skyrmion repulsion,
applied drive, and Magnus forces,
the skyrmions push each other toward
the upper edge of the divot until the topmost
skyrmion escapes from the divot and flows to the next one.
This results in a chain-like motion.
When $F^D=2.5$, the motion ceases due to
the increased drive and Magnus forces, so
the skyrmions accumulate in the divots, producing a reentrant pinning phase.
The location of this reentrant pinned state is shown
in Fig.~\ref{fig4}(c).
Reentrant pinning only occurs for $\alpha_m/\alpha_d>0.2$, while
for $\alpha_m/\alpha_d<0.2$, the
Magnus force is not strong enough to trap the skyrmions
inside the divots. 
We expect that for drives higher than those considered here,
skyrmions can still become pinned inside divots
at smaller $\alpha_m/\alpha_d$.
For $0.2 < \alpha_m/\alpha_d < 1.0$,
the onset of the reentrant pinning phase shifts to lower
applied drives with increasing $\alpha_m/\alpha_d$, and the
threshold then stabilizes at $F^D=2.36$ over the range
$1.0 < \alpha_m/\alpha_d < 2.75$.
The threshold drops again with increasing $\alpha_m/\alpha_d$ above
$\alpha_m/\alpha_d=2.75$.
We find that 
the Magnus force plays a major role
in producing the reentrant pinning.
The high asymmetry system S1 could serve as a topological sorter, since 
skyrmions with higher Magnus terms would be guided into the divots,
while
skyrmions with
reduced Magnus terms can flow without entering the divots.
This is in addition to
other recent proposals for how to sort skyrmions
\cite{vizarim_skyrmion_2020,vizarim_directional_2021,song_guiding_2020}.
The ability to switch between skyrmion motion and
reentrant pinning with an external drive 
can also be used to transfer data:
for low drives, the skyrmions are able to flow and transport information, while
for higher drives, they
become pinned in the divots.

\section{System S2 with lower asymmetry}

\begin{figure}[h]
\centering
\includegraphics[width=0.32\columnwidth]{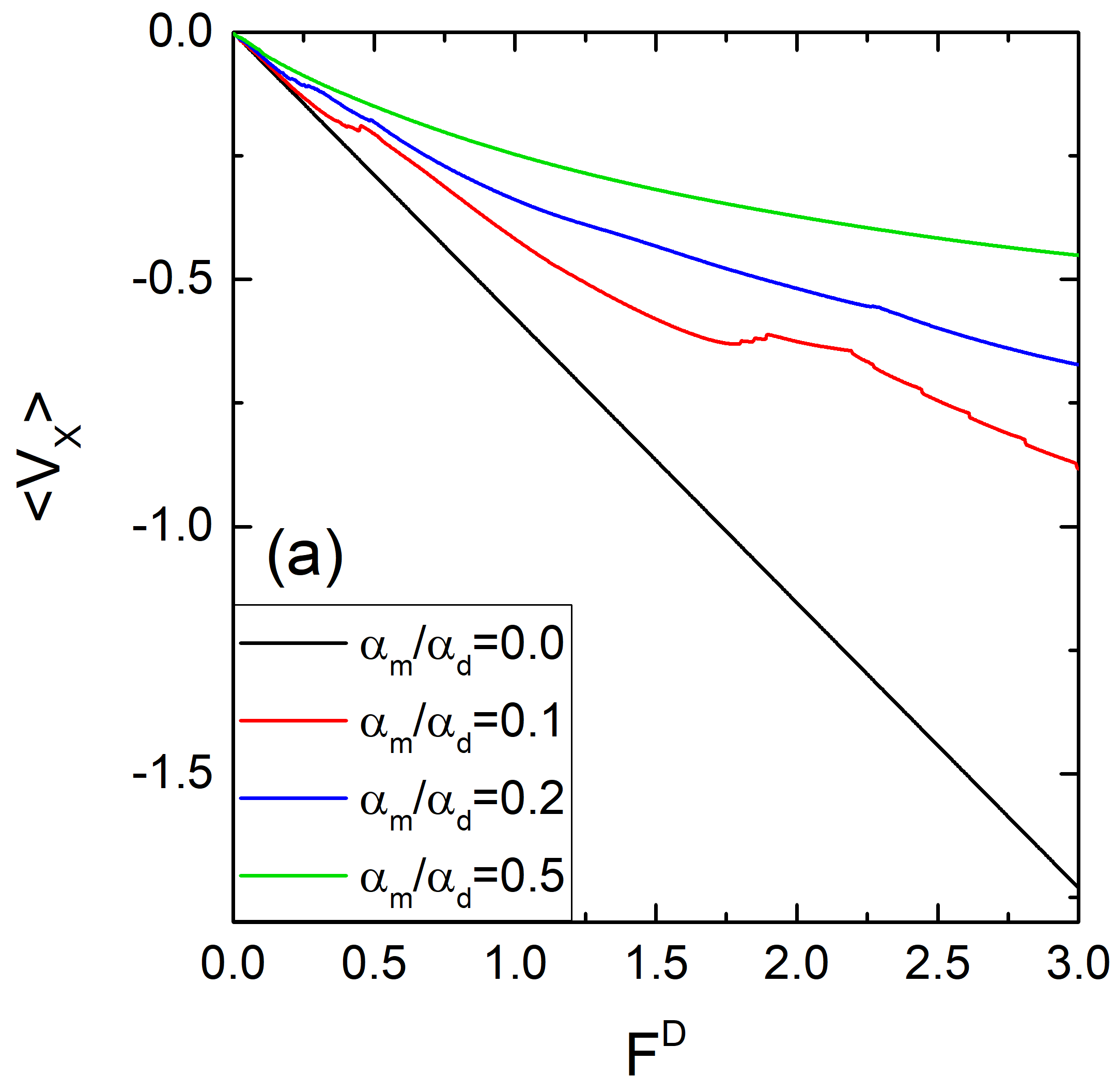}
\includegraphics[width=0.32\columnwidth]{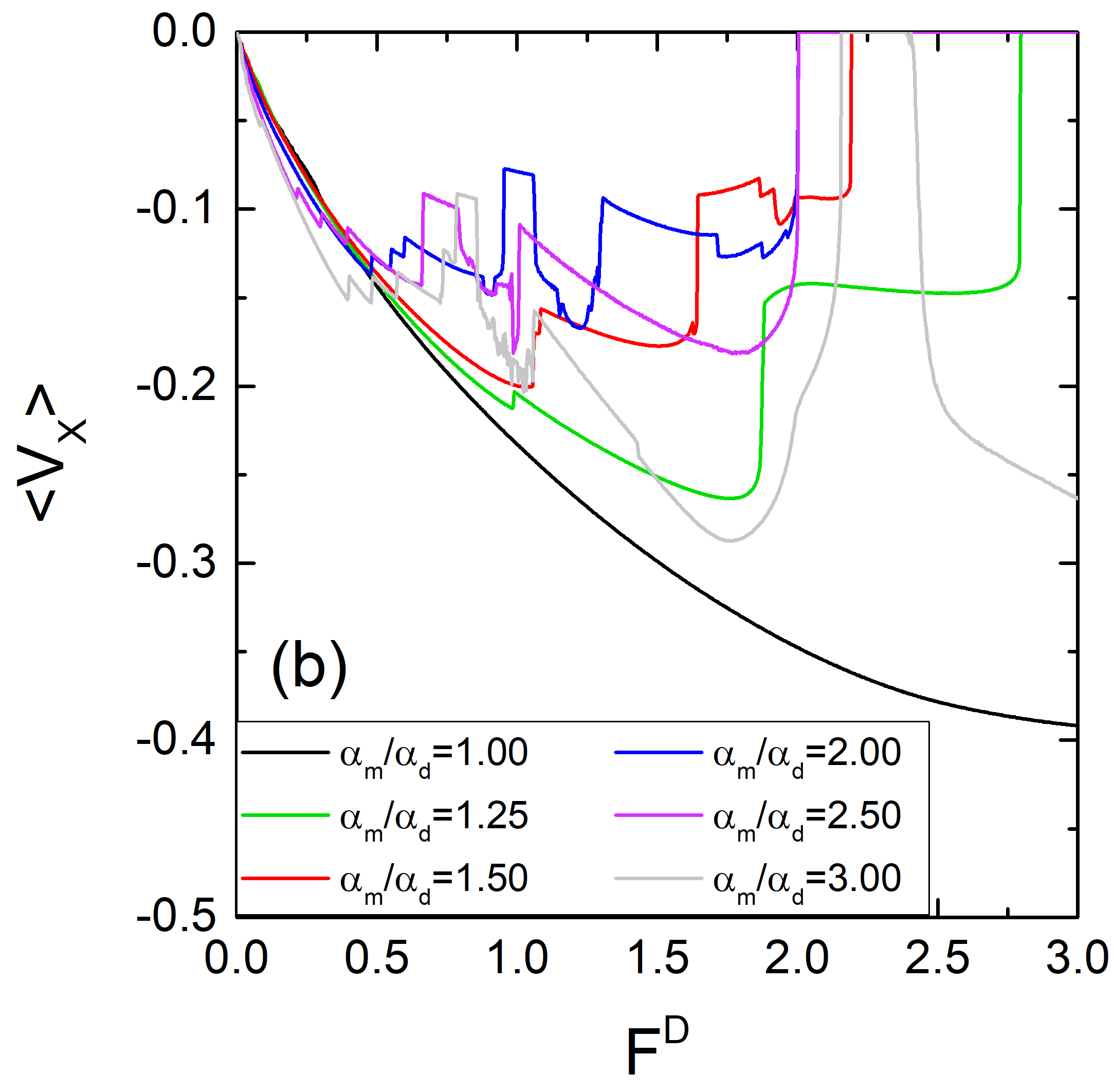}
\includegraphics[width=0.32\columnwidth]{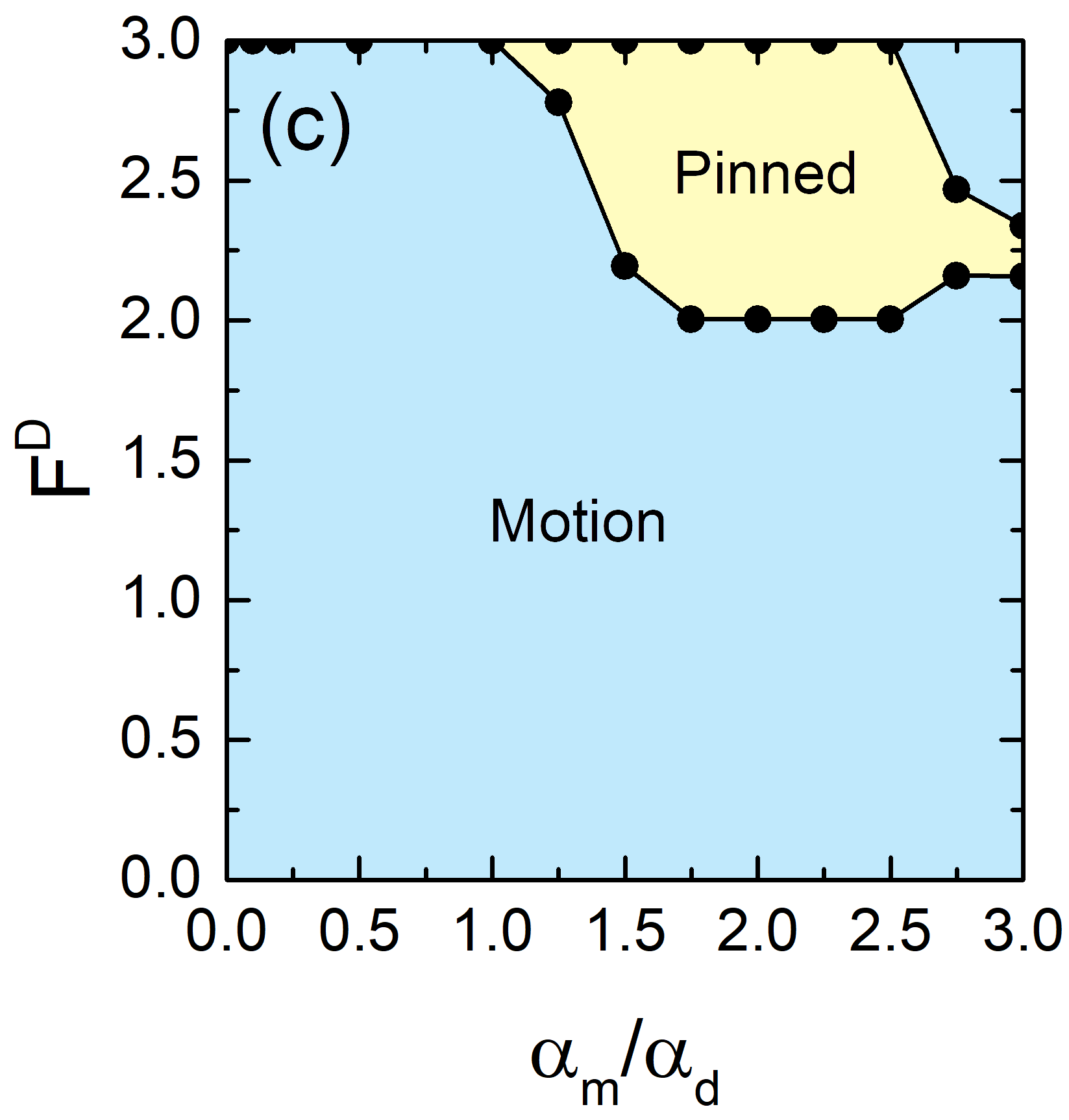}
\caption{Results for easy direction driving along ${\bf \hat d}=-{\bf \hat x}$
for the lower asymmetry system S2.
(a) $\left\langle V_x\right\rangle$ vs $F^D$
for several values of $\alpha_m/\alpha_d$
in the weak Magnus force regime,  $0 \leq \alpha_m/\alpha_d \leq 0.5$.
(b) The same for the intermediate Magnus force
regime, $1.0 \leq \alpha_m/\alpha_d \leq 3.0$.
(c) Dynamical phase diagram as a function
of $F^D$ vs $\alpha_m/\alpha_d$ showing the reentrant pinned phase (yellow)
and the moving phase (blue).}
    \label{fig6}
\end{figure}

We next consider system S2, illustrated
in Fig.~\ref{fig1}(b), where the asymmetry
between the upper and lower part of the sample is reduced.
Note that since the lower part of system S2 is exactly the same as that
of system S1,
the behavior under ${\bf \hat d}=+{\bf \hat x}$
or hard direction driving is identical
for both systems and is as described in the previous section.
Thus, in this section we only consider easy direction
driving with ${\bf \hat d}=-{\bf \hat x}$.
For $-x$
driving, the Magnus force
pushes the skyrmions towards the upper part of the sample, 
where they interact with the shallow upper divots.
An overview of $\left\langle V_x\right\rangle$ versus $F^D$
in the lower asymmetry system S2 for different values of
$\alpha_m/\alpha_d$ appears in Fig.~\ref{fig6}.

\begin{figure}[h]
\centering
\includegraphics[width=0.32\columnwidth]{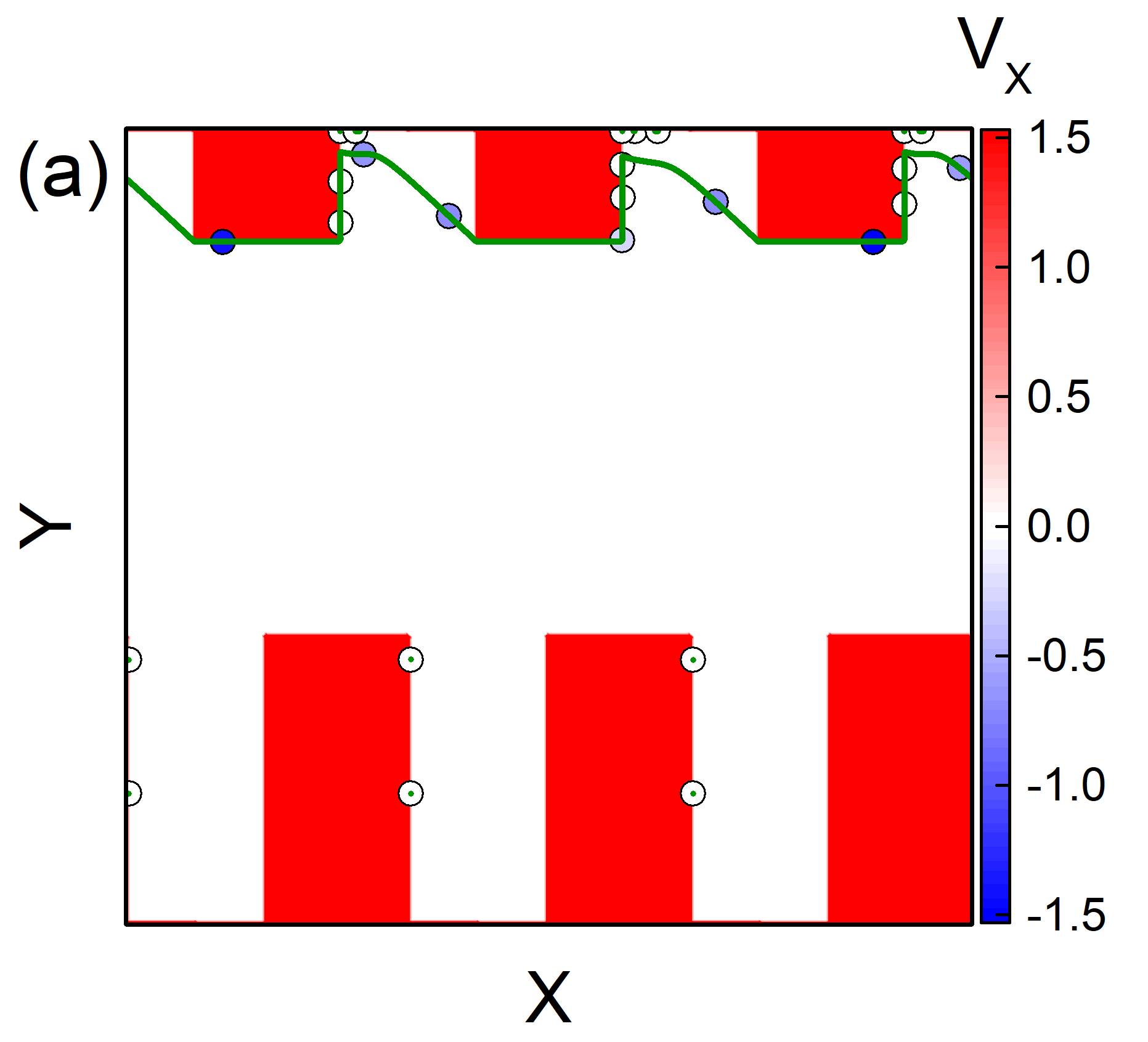}
\includegraphics[width=0.32\columnwidth]{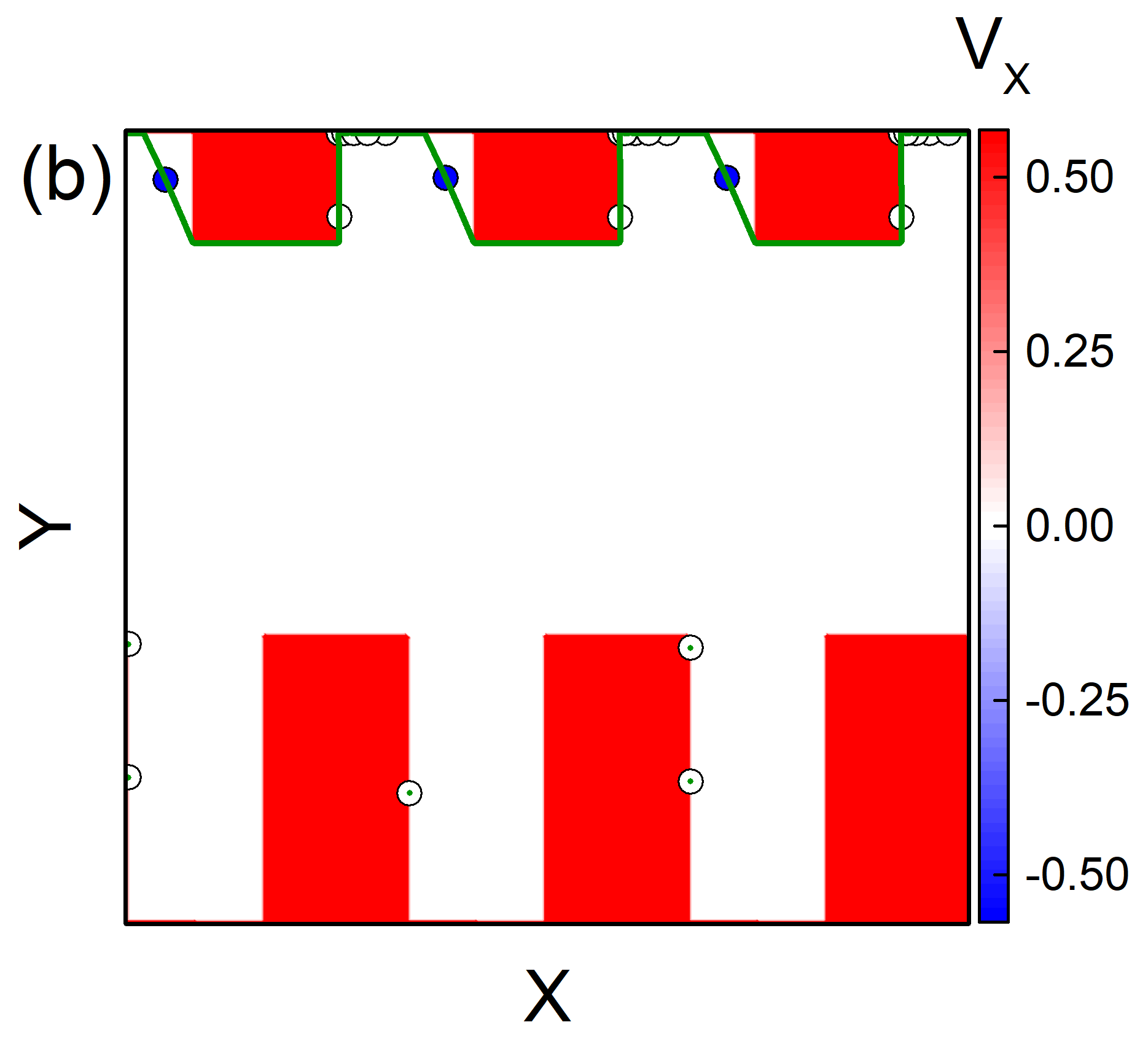}
\includegraphics[width=0.32\columnwidth]{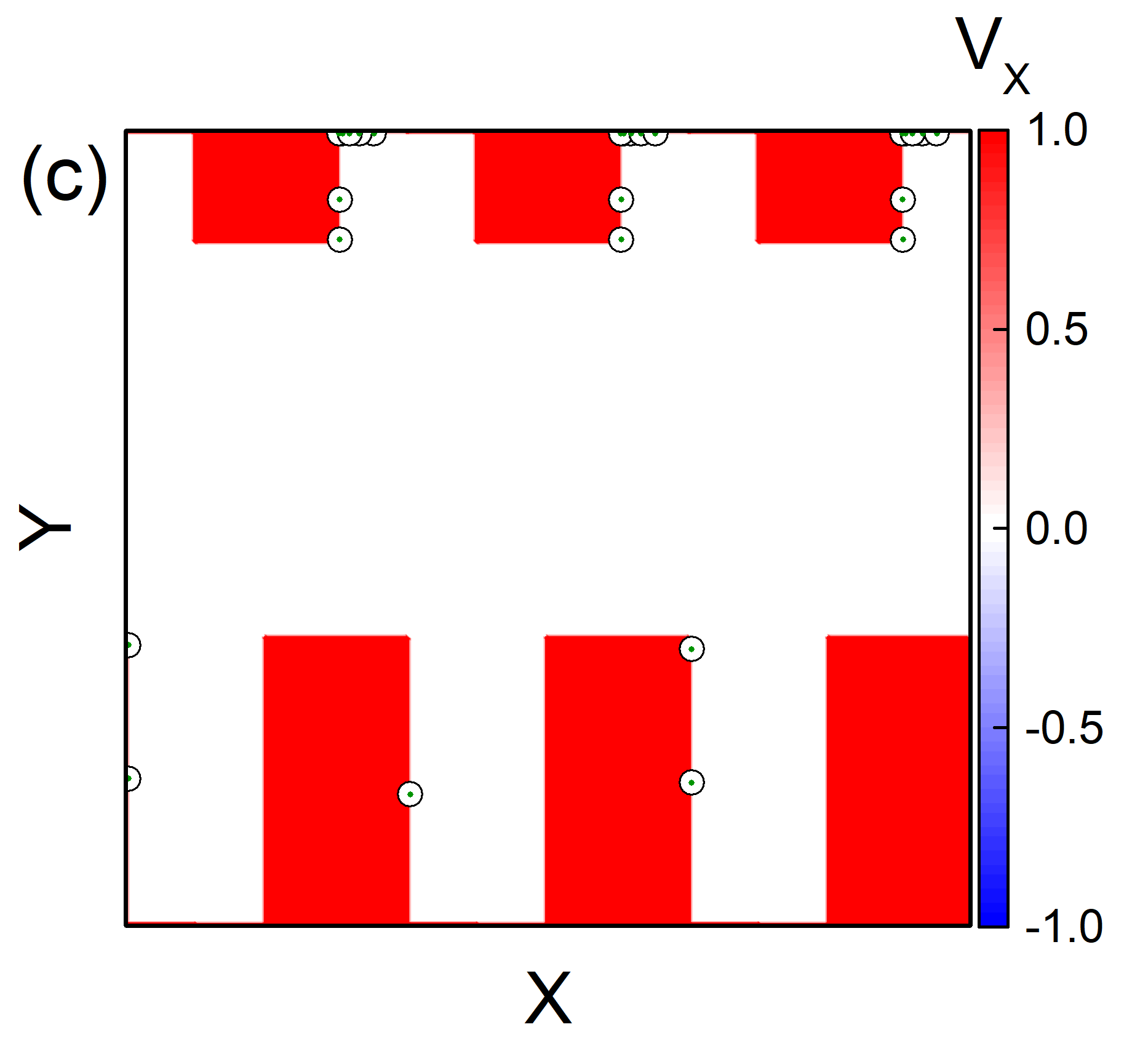}
\caption{Skyrmion positions (black dots) and trajectories (green lines) for
${\bf \hat d}=-{\bf \hat x}$ driving in the lower asymmetry S2 system.
Red areas are forbidden to the skyrmions.
Each skyrmion instantaneous velocity, $V_x$, 
is represented by a color scale attached to the plot, where
white are skyrmions with null velocity, blue negative and red positive velocities. 
(a) At $\alpha_m/\alpha_d=1.0$ and $F^D=1.0$, only a portion of the
skyrmions in the upper portion of the sample are flowing, while the
other skyrmions remain pinned.
(b) At $\alpha_m/\alpha_d=2.5$ and
$F^D=1.5$, all of the skyrmions in the upper part of the sample 
are moving.
(c) At $\alpha_m/\alpha_d=2.5$ and $F^D=2.5$,
the system is in the reentrant pinning phase.
}
    \label{fig7}
\end{figure}

In the low Magnus force regime
shown in Fig.~\ref{fig6}(a), we again find that
when $\alpha_m/\alpha_d=0$, the skyrmion average velocity 
increases linearly with
increasing $F^D$ due to the absence of a Magnus force.
When $\alpha_m/\alpha_d>0$, however,
the $\langle V_x \rangle$ curves become non-linear and the
skyrmion velocity diminishes with
increasing $\alpha_m/\alpha_d$.
In addition, when $\alpha_m/\alpha_d$ becomes large enough,
a reentrant pinned phase emerges as shown in 
Fig.~\ref{fig6}(b).
Many of the features
shown in Fig.~\ref{fig6}(a,b) are similar to those
described in previous section. This is due to the fact that,
apart from a sign change in the driving direction,
the dynamics of the skyrmions in
Fig.~\ref{fig4}(a,b) and in Fig.~\ref{fig6}(a,b)
differs only in the depth of the divots.
The deeper divots can more easily pin the skyrmions,
while the shallower divots permit the emergence of
more complex dynamic regimes, 
in which transitions between pinned states and moving phases
can occur under fine adjustments of the external drive.
In Fig.~\ref{fig6}(c) we plot a dynamic phase diagram
for easy direction driving in the lower asymmetry system S2
as a function of $F^D$ versus $\alpha_m/\alpha_d$.
The extent of the reentrant pinned phase is 
reduced compared to what is shown in Fig.~\ref{fig4}(c)
due to the reduced depth of the divots.
Here, skyrmions that have been
trapped in the divots can begin to flow again when the drive
increases,
as shown for
the samples with $\alpha_m/\alpha_d>2.5$.
Thus, the depth of the divots is another important parameter
that can determine the response of a device constructed using a geometry
of this type.
In Fig.~\ref{fig7}(a), the snapshot of the skyrmion trajectories at
$\alpha_m/\alpha_d=1.0$
shows that not all of the skyrmions in the upper part of the
sample are flowing. Instead, some skyrmions remain trapped while the other
skyrmions flow. 
When $\alpha_m/\alpha_d=2.5$, Fig.~\ref{fig7}(b) shows that
at $F^D=1.5$, all of the skyrmions in the upper part of the sample
are moving; however, when the drive is increased to $F^D=2.5$,
the system enters the reentrant pinned phase, illustrated in
Fig.~\ref{fig7}(c).

\section{The influence of skyrmion density}

\begin{figure}[h]
\centering
\includegraphics[width=0.4\columnwidth]{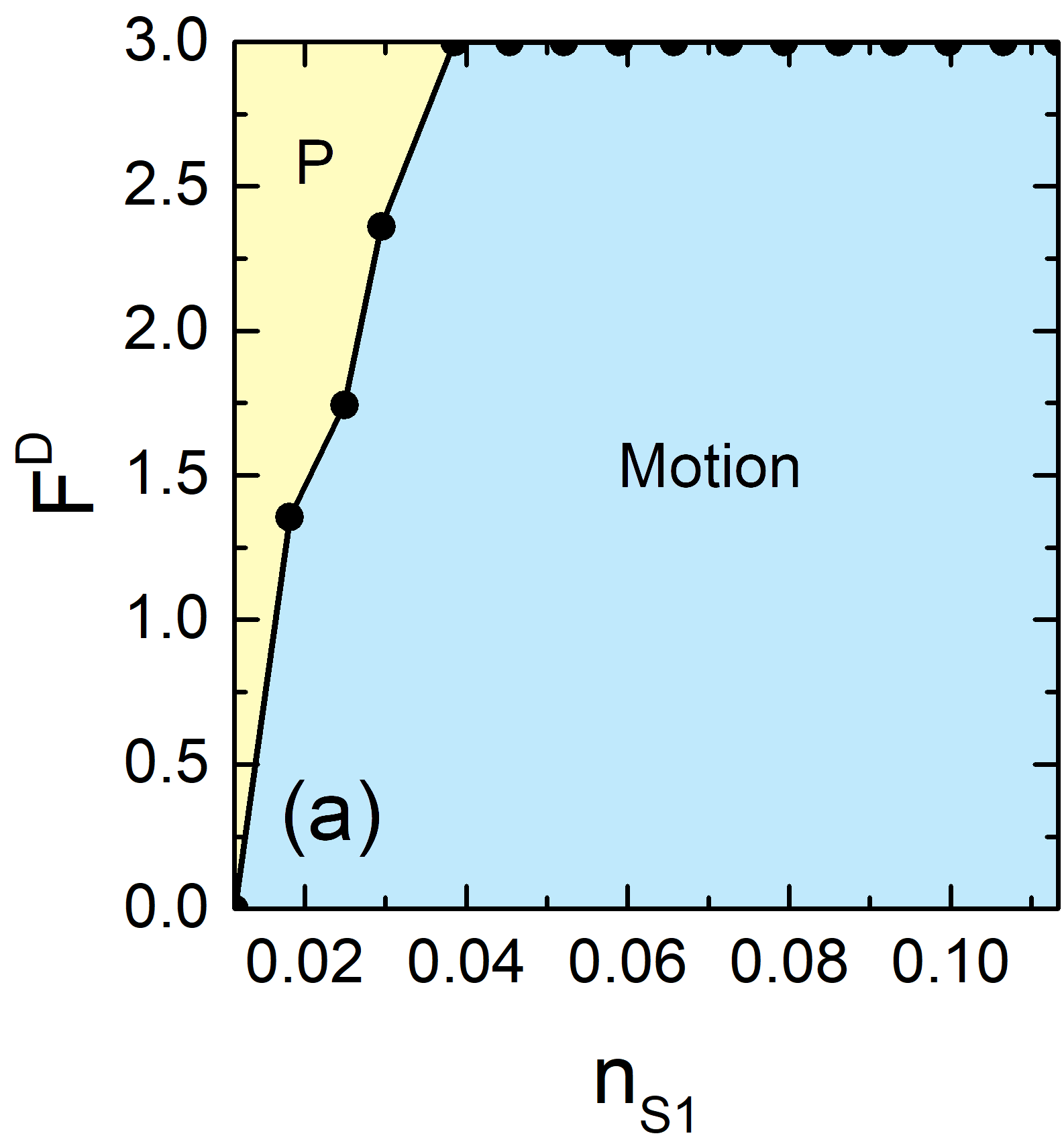}
\includegraphics[width=0.415\columnwidth]{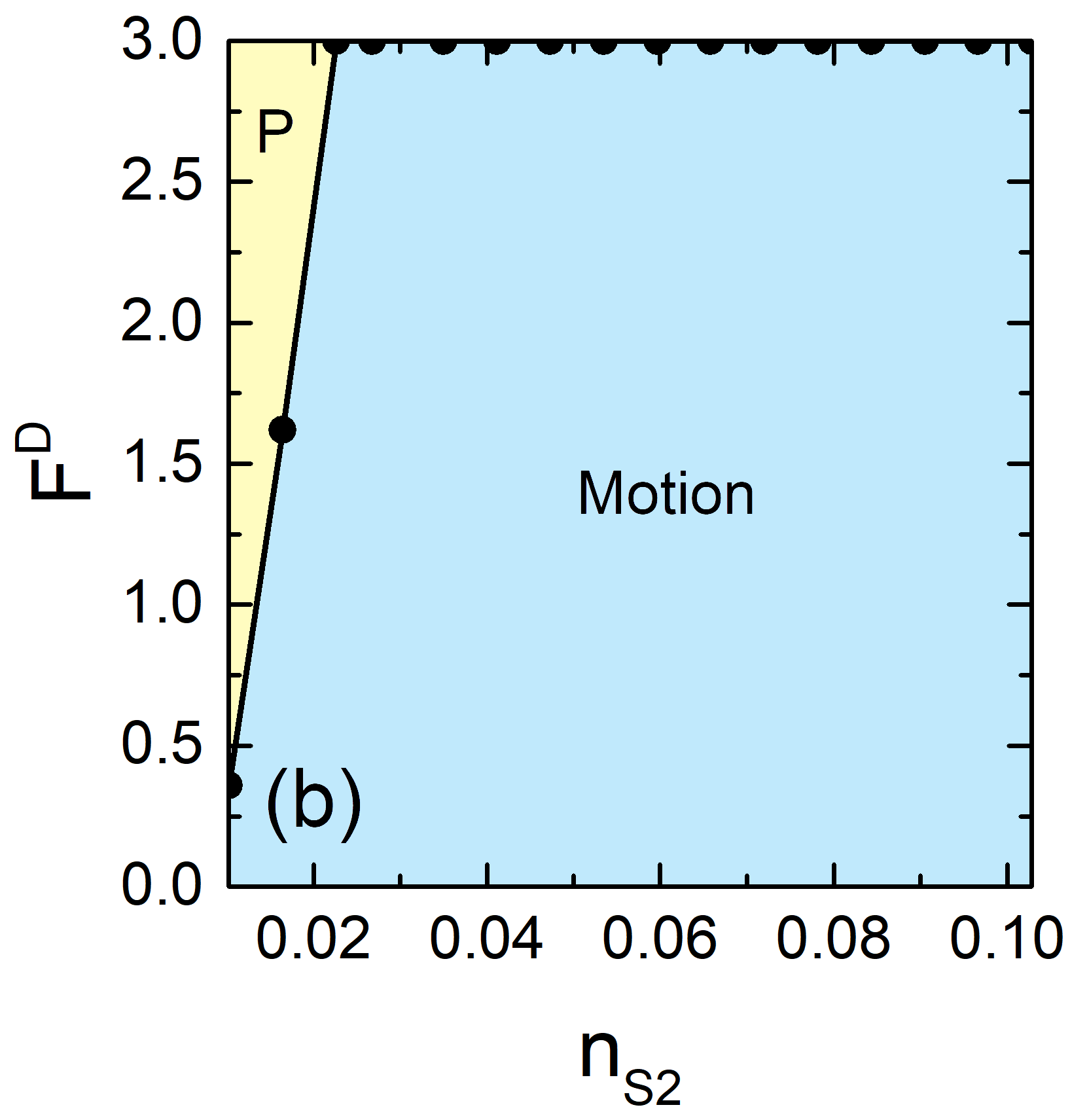}
\includegraphics[width=0.4\columnwidth]{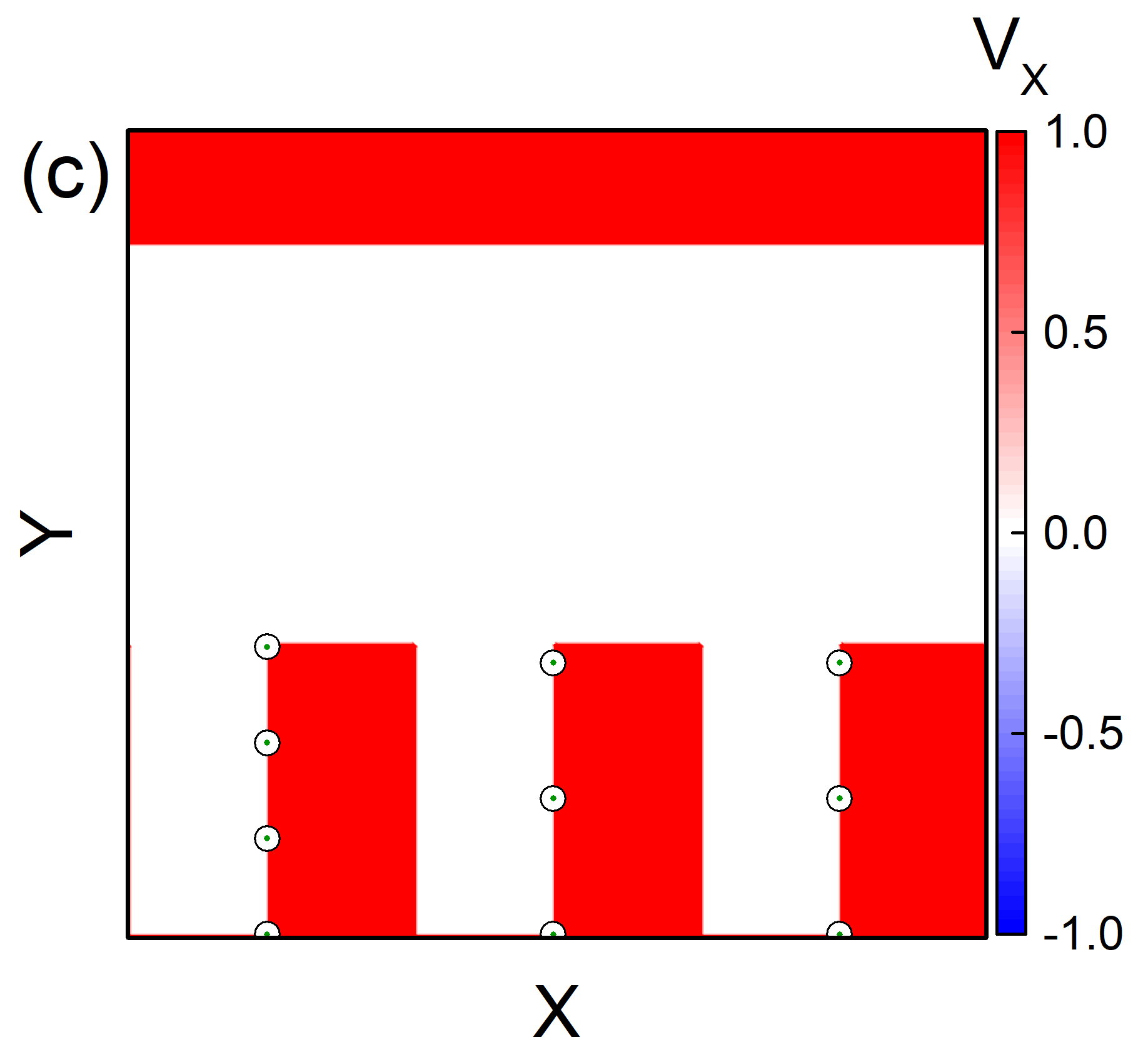}
\includegraphics[width=0.4\columnwidth]{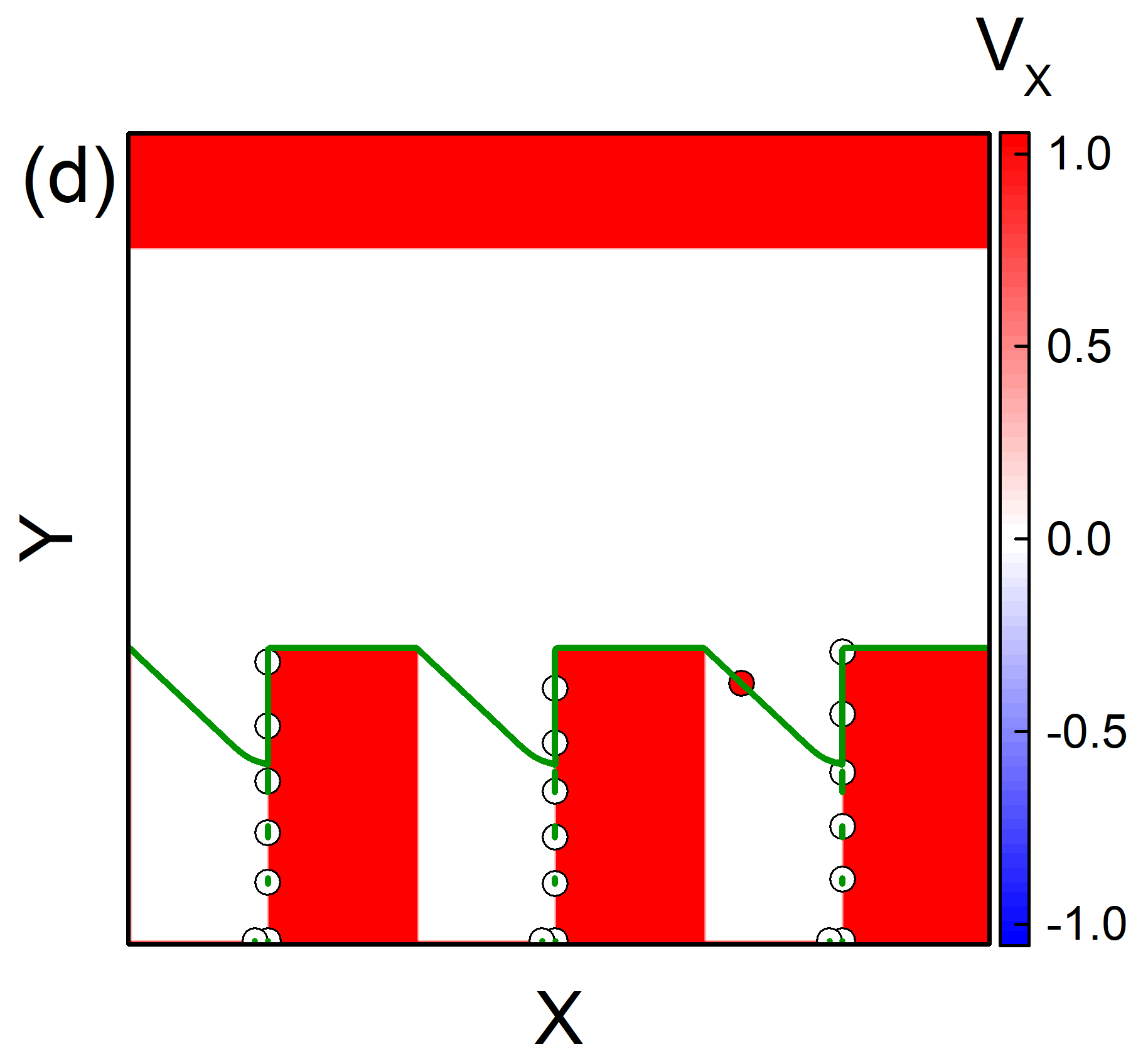}
\caption{(a,b) Dynamical phase diagrams
as a function of $F^D$ vs skyrmion density
(a) $n_{S1}$
in high asymmetry system S1 for hard direction driving
with ${\bf \hat d}=-{\bf \hat x}$ and
(b) $n_{S2}$ in lower asymmetry system S2 for easy direction driving
with ${\bf \hat d}=+{\bf \hat x}$ for samples with
$\alpha_m/\alpha_d=1.0$.
Moving regions are shown in blue and pinned (P) regions are shown in yellow.
(c,d) Skyrmion positions (black dots) and trajectories (green lines)
for the high asymmetry system S1 under hard direction driving
with ${\bf \hat d}=-{\bf \hat x}$.
Red areas are forbidden to the skyrmions.
Each skyrmion velocity is represented by a color scale attached to the plot, where
white are skyrmions with null velocity, blue negative and red positive velocities.
(c) A pinned state at low $n_{S1}$.
(d) A chain of moving skyrmions for higher $n_{S1}$.
}
    \label{fig8}
\end{figure}

Next we explore how the dynamics change as
a function of skyrmion density.
We vary the number of skyrmions inside
the sample between $N_{sk}=10$ and $N_{sk}=100$ while holding
the Magnus intensity fixed at $\alpha_m/\alpha_d = 1.0$.
In Fig.~\ref{fig8}(a) we plot a dynamic phase diagram as a function
of $F^D$ versus skyrmion density for the high asymmetry system S1 under
hard direction driving with ${\bf \hat d}=-{\bf \hat x}$, and
in Fig.~\ref{fig8}(b) we show the dynamic phases for the lower asymmetry
system S2 under
easy direction driving with ${\bf \hat d}=+{\bf \hat x}$.
As a function of increasing skyrmion density or decreasing drive, there is
a transition in both cases from the pinned to the moving state.
The pinned phase is only present for low skyrmion densities,
indicating that diode behavior would be most prominent at
densities where
the skyrmions are widely spaced and their motion is dominated
by the skyrmion Hall angle.
In this situation, the skyrmions are driven against the divots
where they become trapped, as illustrated in
Fig.~\ref{fig8}(c).
When the skyrmion density increases,
the skyrmions approach each other more closely,
and a competition between
the skyrmion-skyrmion
interactions and the skyrmion Hall effect emerges.
This leads to a chain motion
of a portion of the skyrmions that never become
stuck in the divots, as shown in Fig.~\ref{fig8}(d).
Furthermore,
in sample S1 with driving along
${\bf \hat d}=+{\bf \hat x}$, shown in Fig.~\ref{fig8}(a), 
the skyrmions experience the influence of the deeper divots,
resulting in the presence of
pinned phases for $n_{S1} \leq 0.0385$.
On the other hand, in sample S2 for driving along
${\bf \hat d}=-{\bf \hat x}$, shown in Fig.~\ref{fig8}(b), 
the skyrmions interact with the shallower upper divots,
resulting in a pinned phase
only over the reduced range $n_{S2} \leq 0.0226$.
These results indicate that introducing
deeper divots could produce stronger pinning effects for
the skyrmions.
In order to make a diode device that can support a high
skyrmion density,
a good option
would be to use the high asymmetry geometry S1 with deeper divots.
This would result in an easy flow for driving in one direction and
a pinned state or hindered motion for driving in the other direction.
The depth of the divots is
an important parameter for controlling the skyrmion motion.

\section{Diode Effect}

In the previous sections we showed that skyrmion transport in
both samples S1 and S2 exhibits a very significant
dependence on the direction of the applied drive.
In the high asymmetry sample S1, when the drive is applied along the 
easy direction, the velocity of the skyrmions increases linearly with
increasing drive.
On the other hand, when the drive is applied along the
hard direction, the skyrmions may get stuck, and their velocity
is strongly reduced.
We also showed that in the lower asymmetry sample S2,
the direction of the drive is very important in determining the nature of
the skyrmion motion. 
Although in this case the asymmetry is reduced,
it is also possible to see clear differences in the
skyrmion motion depending on the direction of the drive.

\begin{figure}[h]
\centering
\includegraphics[width=0.4\columnwidth]{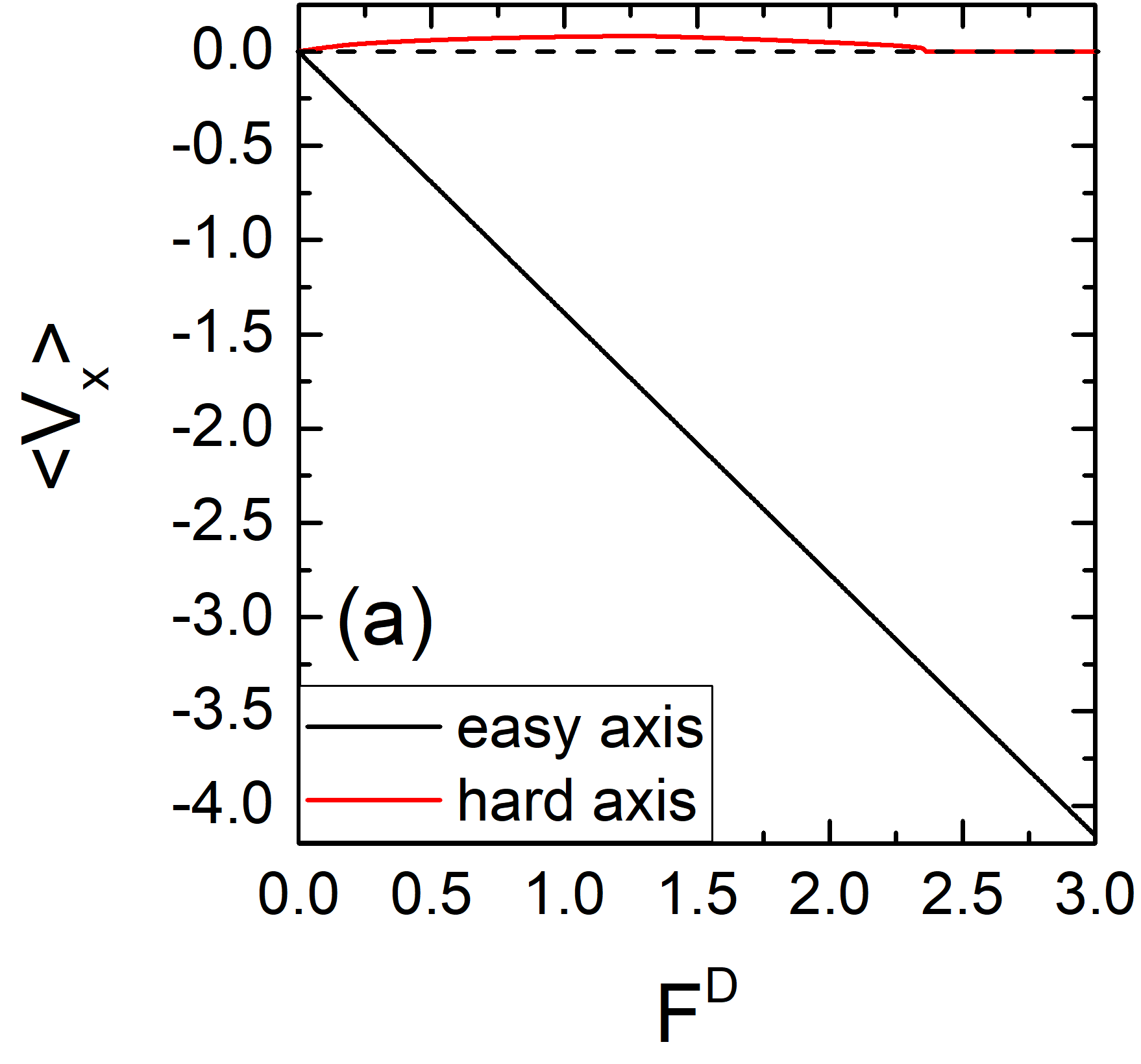}
\includegraphics[width=0.4\columnwidth]{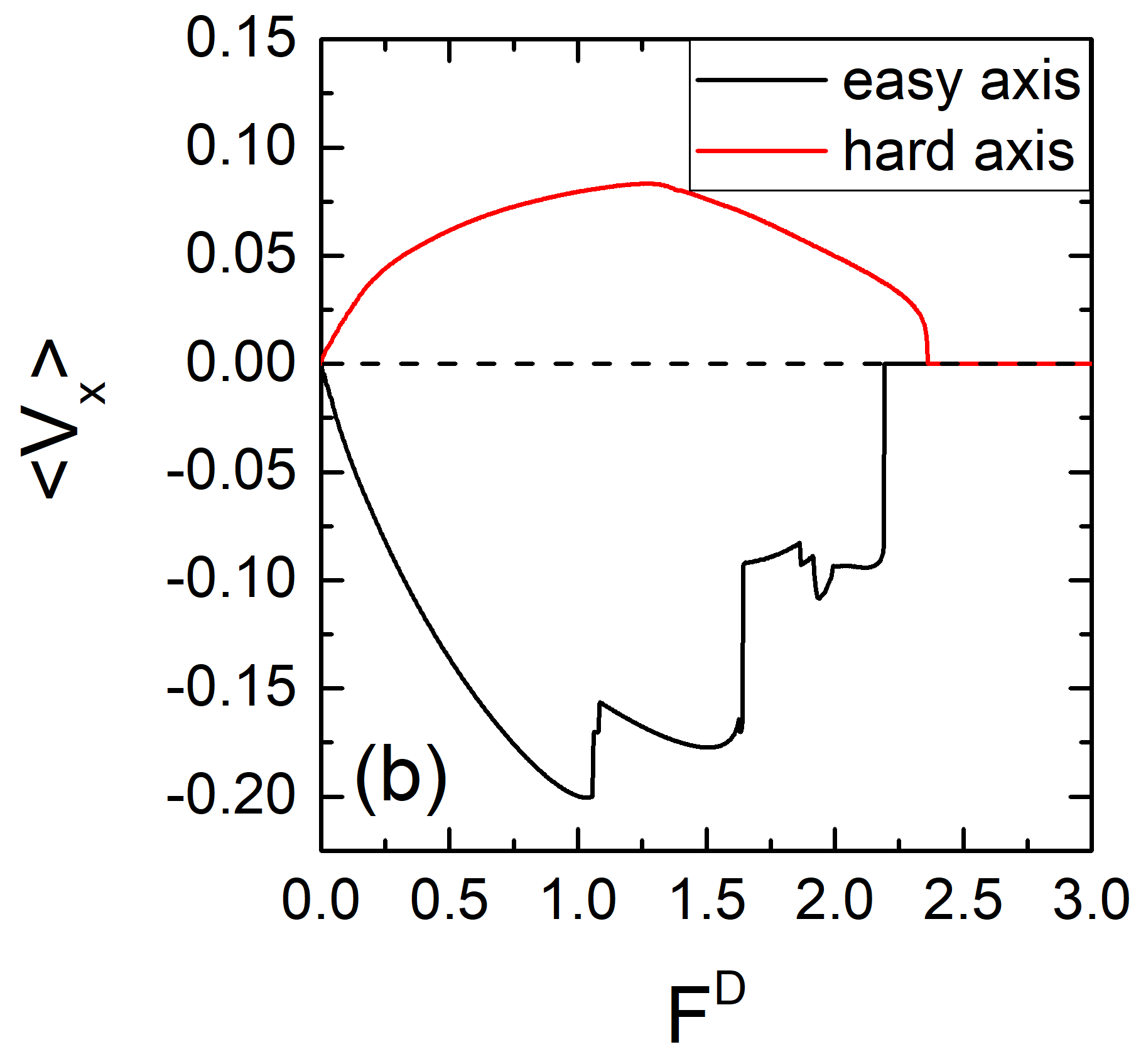}
\caption{$\left\langle V_x\right\rangle$ vs $F^D$
for samples with $\alpha_m/\alpha_d=1.5$
under easy ${\bf \hat d}=+{\bf \hat x}$ (black) and
hard ${\bf \hat d}=+{\bf \hat x}$ (red) direction driving.
(a) High asymmetry sample S1.
(b) Lower asymmetry sample S2.
The dashed line indicates in both figures where $\left\langle V_x\right\rangle = 0$.
}
\label{fig9}
\end{figure}

In Fig.~\ref{fig9} we show a comparison
of $\langle V_x\rangle$ versus $F^D$ for
driving in the easy and hard
directions for samples S1 and S2 at $\alpha_m/\alpha_d=1.5$.
In the high asymmetry sample S1, illustrated in Fig.~\ref{fig1}(a),
changing the driving direction produces very different behaviors.
Skyrmions driven along the ${\bf \hat d}=-{\bf \hat x}$ easy direction
have a velocity that increases in magnitude with increasing drive.
For driving along the
${\bf \hat d}=+{\bf \hat x}$ hard direction,
the magnitude of the skyrmion velocity
is greatly reduced.
Additionally, $\langle V_x\rangle$ drops to zero
when $F^D>2.36$, so for drives in this window, a device
could be created that gives
insulating transport behavior of the skyrmions for motion in one
direction and conducting transport behavior for motion in the
other direction,
similar to what was proposed in Refs.~\cite{Jung21} and \cite{Feng22}.
The Magnus term pushes the skyrmions toward the divots when the
drive is in the hard direction,
and the skyrmions must overcome the potential barrier of the divots
via the chain motion
described in 
previous sections. 

In Fig.~\ref{fig9}(b), the $\langle V_x\rangle$ versus $F^D$ curves
for easy and hard direction driving in the lower asymmetry system S2 indicate
that the reduction of asymmetry produces significant changes in the
behavior.
The magnitude of the skyrmion velocity remains higher for easy
direction than for hard direction driving due to the reduced depths of the
divots traversed by the skyrmions during easy direction driving.
Thus, the depth of the divots is a very important parameter that can control
the skyrmion velocity.
Additionally, for both directions of drive,
the system enters a reentrant pinning phase,
although the threshold drive above which this phase appears is lower
for easy direction than for hard direction driving.
These features make it possible to construct different types of diodes
using the high asymmetry sample S1 and the low asymmetry sample S2.

\section{Discussion}

The asymmetric potentials proposed in this work
generate a diode motion of skyrmions
when the external driving force is applied in one direction or the other.
The diode effect emerges from the combination
of the asymmetric potentials and the skyrmion Hall effect.
The presence of the Hall effect is crucial,
and we demonstrated that skyrmions with no Magnus term,
such as antiferromagnetic skyrmions
\cite{barker_static_2016,legrand_room-temperature_2020}
or liquid crystal skyrmions \cite{duzgun_commensurate_2020},
do not exhibit a diode effect.
Moreover, we showed that both strong and weak
diodes can be constructed.
Since we consider asymmetric potentials, if
ac rather than dc driving were applied,
we expect that clear ratchet effects would emerge in which there
would be a net motion of skyrmions along the easy direction.
There have been a number of proposals for how to construct
ratchet geometries for skyrmions,
such as by using asymmetric potentials
\cite{reichhardt_magnus-induced_2015,ma_reversible_2017, souza_skyrmion_2021},
interfaces \cite{vizarim_guided_2021,Velez22},
or even purely periodic obstacle lattices \cite{gobel_skyrmion_2021}.
We note that for asymmetric substrates
such as sawtooth potentials, a ratchet or diode effect
will still occur even in the absence of a Magnus force.

In the model considered in this work, the
skyrmions are rigid bodies that can not deform, merge,
or be created or annihilated.
It would be interesting to investigate multiple skyrmions
in asymmetric potentials of the types we consider
using continuum-based simulations, where the impact of skyrmion
deformations could be explored.
Our results should be valid in the limit of low skyrmion densities and
low currents, where internal modes of the skyrmions are
not excited.

With regard to thermal effects,
it is known that temperature can change
the location of transition points or even 
induce skyrmion creep \cite{reichhardt_thermal_2018,vizarim_skyrmion_2020}. 
We expect that inclusion of a
finite temperatures would modify the reentrant pinned phases;
however, for low temperatures we expect our results to be accurate. 
One possible way to mitigate thermal effects on the diode behavior
is to increase the depth of the divots,
making the trapping of skyrmions more efficient.

We also note that our proposed geometries
could be used for sorting mixtures of skyrmions
with different skyrmion Hall angles, where the
skyrmions with higher skyrmion Hall angles
would be trapped or move more slowly.
Finally, we note that our geometries can be applied more
generally
to particle-like systems that have a Hall angle,
including active spinners or chiral active matter \cite{Banerjee17},
charges in a magnetic field such as Wigner crystals \cite{Reichhardt21},
and dusty plasmas in magnetic fields \cite{Melzer21}.

\section{Summary}

We have investigated a Magnus-induced diode effect
using a particle based model for skyrmions
in two different channel geometries containing
periodic array of divots on one side of the channel and either
a smooth wall or shallower divots on the other side of the channel.
Under an applied drive, a diode effect emerges since
the skyrmions are deflected to opposite sides of the channel depending
on the driving direction.
When the skyrmions are deflected toward the deeper divots,
they either become completely trapped or exhibit a chain-like flow of
low velocity compared to the motion for driving in the opposite
direction.
We highlight the difference in transport properties for the two
channel geometries 
and show how the results change as the Magnus term increases in importance.
The diode effect we observe only occurs when the Magnus force
is finite. Our geometry
is distinct from
previously considered periodic asymmetric sawtooth potentials,
which produce diode effects even in the absence of a Magnus force.
When multiple interacting skyrmions are present in the channel,
there can be  a variety of dips or jumps in the transport curves
due to collective trapping or, in some cases, escape over the barriers.
This
leads to regimes of negative differential conductivity
or reentrant pinning in which the skyrmions
can flow at low drives but become strongly pinned at higher dives.
Our results should
be general for other particle-based systems
that have a Hall effect, such as chiral active matter,
Wigner crystals in a magnetic field, and magnetized dusty plasma systems.

\ack
This work was supported by the US Department of Energy through the Los Alamos National Laboratory. Los
Alamos National Laboratory is operated by Triad National Security, LLC, for the National Nuclear Security
Administration of the U. S. Department of Energy (Contract No. 892333218NCA000001). 
J.C.B.S and N.P.V acknowledges funding from Fundação de Amparo à Pesquisa do Estado
de São Paulo - FAPESP (Grants 2021/04941-0 and 2017/20976-3 respectively).

\section*{References}

\bibliographystyle{iopart-num}
\bibliography{refs}

\providecommand{\newblock}{}
\begin{thebibliography}{10}
\expandafter\ifx\csname url\endcsname\relax
  \def\url#1{{\tt #1}}\fi
\expandafter\ifx\csname urlprefix\endcsname\relax\def\urlprefix{URL }\fi
\providecommand{\eprint}[2][]{\url{#2}}

\bibitem{Kitai11}
Kitai A 2011 {\em Principles of Solar Cells, LEDs and Diodes\/} (Wiley,
  Chichester)

\bibitem{tocci_thinfilm_1995}
Tocci M~D, Bloemer M~J, Scalora M, Dowling J~P and Bowden C~M 1995 {\em Appl.
  Phys. Lett.\/} {\bf 66} 2324--2326 ISSN 0003-6951

\bibitem{scalora_photonic_1994}
Scalora M, Dowling J~P, Bowden C~M and Bloemer M~J 1994 {\em J. Appl. Phys.\/}
  {\bf 76} 2023--2026

\bibitem{wang_optical_2013}
Wang D~W, Zhou H~T, Guo M~J, Zhang J~X, Evers J and Zhu S~Y 2013 {\em Phys.
  Rev. Lett.\/} {\bf 110} 093901

\bibitem{sciamanna_physics_2015}
Sciamanna M and Shore K~A 2015 {\em Nature Photonics\/} {\bf 9} 151--162 ISSN
  1749-4893

\bibitem{li_thermal_2004}
Li B, Wang L and Casati G 2004 {\em Phys. Rev. Lett.\/} {\bf 93} 184301

\bibitem{martinez-perez_rectification_2015}
Mart{\' \i}nez-P{\' e}rez M~J, Fornieri A and Giazotto F 2015 {\em Nature
  Nanotech.\/} {\bf 10} 303--307

\bibitem{mates_fluid_2014}
Mates J~E, Schutzius T~M, Qin J, Waldroup D~E and Megaridis C~M 2014 {\em ACS
  Appl. Mater. Interfaces\/} {\bf 6} 12837--12843 ISSN 1944-8244

\bibitem{shou_all_2018}
Shou D and Fan J 2018 {\em Adv. Funct. Mater.\/} {\bf 28} 1800269 ISSN
  1616-3028

\bibitem{zhao_ferromagnetic_2020}
Zhao L, Liang X, Xia J, Zhao G and Zhou Y 2020 {\em Nanoscale\/} {\bf 12}
  9507--9516

\bibitem{wang_magnetic_2020}
Wang J, Xia J, Zhang X, Zheng X, Li G, Chen L, Zhou Y, Wu J, Yin H, Chantrell R
  and Xu Y 2020 {\em Appl. Phys. Lett.\/} {\bf 117} 202401

\bibitem{tulapurkar_spin-torque_2005}
Tulapurkar A~A, Suzuki Y, Fukushima A, Kubota H, Maehara H, Tsunekawa K,
  Djayaprawira D~D, Watanabe N and Yuasa S 2005 {\em Nature (London)\/} {\bf
  438} 339--342

\bibitem{song_spin-wave_2021}
Song L, Yang H, Liu B, Meng H, Cao Y and Yan P 2021 {\em J. Mag. Mag. Mater.\/}
  {\bf 532} 167975

\bibitem{Jung21}
Jung D~H, Han H~S, Kim N, Kim G, Jeong S, Lee S, Kang M, Im M~Y and Lee K~S
  2021 {\em Phys. Rev. B\/} {\bf 104}(6) L060408

\bibitem{Feng22}
Feng Y, Zhang X, Zhao G and Xiang G 2022 {\em IEEE Trans. Electron Devices\/}
  {\bf 69} 1293

\bibitem{fang_giant_2016}
Fang B, Carpentieri M, Hao X, Jiang H, Katine J~A, Krivorotov I~N, Ocker B,
  Langer J, Wang K~L, Zhang B, Azzerboni B, Amiri P~K, Finocchio G and Zeng Z
  2016 {\em Nature Commun.\/} {\bf 7} 11259

\bibitem{harrington_practical_2009}
Harrington S~A, MacManus-Driscoll J~L and Durrell J~H 2009 {\em Appl. Phys.
  Lett.\/} {\bf 95} 022518

\bibitem{lyu_superconducting_2021}
Lyu Y~Y, Jiang J, Wang Y~L, Xiao Z~L, Dong S, Chen Q~H, Milo{\v s}evi{\' c}
  M~V, Wang H, Divan R, Pearson J~E, Wu P, Peeters F~M and Kwok W~K 2021 {\em
  Nature Commun.\/} {\bf 12} 2703

\bibitem{lu_reversible_2007}
Lu Q, Reichhardt C~J~O and Reichhardt C 2007 {\em Phys. Rev. B\/} {\bf 75}
  054502

\bibitem{wordenweber_guidance_2004}
W\"ordenweber R, Dymashevski P and Misko V~R 2004 {\em Phys. Rev. B\/} {\bf 69}
  184504

\bibitem{van_de_vondel_vortex-rectification_2005}
Van~de Vondel J, de~Souza~Silva C~C, Zhu B~Y, Morelle M and Moshchalkov V~V
  2005 {\em Phys. Rev. Lett.\/} {\bf 94} 057003

\bibitem{olson_reichhardt_vortex_2013}
Olson~Reichhardt C~J and Reichhardt C 2013 {\em J. Supercond. Novel Mag.\/}
  {\bf 26} 2005--2008

\bibitem{reichhardt_jamming_2010}
Reichhardt C and Olson~Reichhardt C~J 2010 {\em Physica C\/} {\bf 470} 722--725

\bibitem{muhlbauer_skyrmion_2009}
M{\" u}hlbauer S, Binz B, Jonietz F, Pfleiderer C, Rosch A, Neubauer A, Georgii
  R and B{\" o}ni P 2009 {\em Science\/} {\bf 323} 915--919

\bibitem{fert_skyrmions_2013}
Fert A, Cros V and Sampaio J 2013 {\em Nature Nanotechnol.\/} {\bf 8} 152--156

\bibitem{fert_magnetic_2017}
Fert A, Reyren N and Cros V 2017 {\em Nature Rev. Mater.\/} {\bf 2} 1--15

\bibitem{jonietz_spin_2010}
Jonietz F, M{\" u}hlbauer S, Pfleiderer C, Neubauer A, M{\" u}nzer W, Bauer A,
  Adams T, Georgii R, B{\" o}ni P, Duine R~A, Everschor K, Garst M and Rosch A
  2010 {\em Science\/} {\bf 330} 1648--1651

\bibitem{schulz_emergent_2012}
Schulz T, Ritz R, Bauer A, Halder M, Wagner M, Franz C, Pfleiderer C, Everschor
  K, Garst M and Rosch A 2012 {\em Nature Phys.\/} {\bf 8} 301--304

\bibitem{yu_skyrmion_2012}
Yu X~Z, Kanazawa N, Zhang W~Z, Nagai T, Hara T, Kimoto K, Matsui Y, Onose Y and
  Tokura Y 2012 {\em Nature Commun.\/} {\bf 3}

\bibitem{olson_reichhardt_comparing_2014}
Olson~Reichhardt C, Lin S, Ray D and Reichhardt C 2014 {\em Physica C\/} {\bf
  503} 52--57

\bibitem{nagaosa_topological_2013}
Nagaosa N and Tokura Y 2013 {\em Nature Nanotechnol.\/} {\bf 8} 899--911

\bibitem{iwasaki_universal_2013}
Iwasaki J, Mochizuki M and Nagaosa N 2013 {\em Nature Commun.\/} {\bf 4} 1463

\bibitem{jiang_direct_2017}
Jiang W, Zhang X, Yu G, Zhang W, Wang X, Jungfleisch M~B, Pearson J~E, Cheng X,
  Heinonen O, Wang K~L, Zhou Y, Hoffmann A and te~Velthuis S~G~E 2017 {\em
  Nature Phys.\/} {\bf 13} 162--169

\bibitem{zeissler_diameter-independent_2020}
Zeissler K, Finizio S, Barton C, Huxtable A~J, Massey J, Raabe J, Sadovnikov
  A~V, Nikitov S~A, Brearton R, Hesjedal T, van~der Laan G, Rosamond M~C,
  Linfield E~H, Burnell G and Marrows C~H 2020 {\em Nature Commun.\/} {\bf 11}
  428

\bibitem{litzius_skyrmion_2017}
Litzius K, Lemesh I, Kr{\" u}ger B, Bassirian P, Caretta L, Richter K, B{\"
  u}ttner F, Sato K, Tretiakov O~A, F{\" o}rster J, Reeve R~M, Weigand M,
  Bykova I, Stoll H, Sch{\" u}tz G, Beach G~S~D and Kl{\" a}ui M 2017 {\em
  Nature Phys.\/} {\bf 13} 170--175

\bibitem{Shu22}
Shu Y, Li Q, Xia J, Lai P, Hou Z, Zhao Y, Zhang D, Zhou Y, Liu X and Zhao G
  2022 {\em Appl. Phys. Lett.\/} {\bf 121} 042402

\bibitem{reichhardt_magnus-induced_2015}
Reichhardt C, Ray D and Reichhardt C~J~O 2015 {\em New J. Phys.\/} {\bf 17}
  073034

\bibitem{ma_reversible_2017}
Ma X, Reichhardt C~J~O and Reichhardt C 2017 {\em Phys. Rev. B\/} {\bf 95}
  104401

\bibitem{souza_skyrmion_2021}
Souza J~C~B, Vizarim N~P, Reichhardt C~J~O, Reichhardt C and Venegas P~A 2021
  {\em Phys. Rev. B\/} {\bf 104} 054434

\bibitem{gobel_skyrmion_2021}
G{\" o}bel B and Mertig I 2021 {\em Sci. Rep.\/} {\bf 11} 3020

\bibitem{Banerjee17}
Banerjee D, Souslov A, Abanov A~G and Vitelli V 2017 {\em Nature Commun.\/}
  {\bf 8} 1573

\bibitem{Reichhardt21}
Reichhardt C and Reichhardt C~J~O 2021 {\em Phys. Rev. B\/} {\bf 103}(12)
  125107

\bibitem{Melzer21}
Melzer A, Kr{\" u}ger H, Maier D and Sch{\" u}tt S 2021 {\em Rev. Mod. Plasma
  Phys.\/} {\bf 5} 11

\bibitem{Lin13}
Lin S~Z, Reichhardt C, Batista C~D and Saxena A 2013 {\em Phys. Rev. B\/} {\bf
  87}(21) 214419

\bibitem{souza_clogging_2022}
Souza J~C~B, Vizarim N~P, Reichhardt C~J~O, Reichhardt C and Venegas P~A 2022
  Clogging, diode and collective effects of skyrmions in funnel geometries

\bibitem{vizarim_skyrmion_2020}
Vizarim N~P, Reichhardt C~J~O, Venegas P~A and Reichhardt C 2020 {\em J. Phys.
  Commun.\/} {\bf 4} 085001

\bibitem{vizarim_directional_2021}
Vizarim N~P, Souza J~C~B, Reichhardt C, Reichhardt C~J~O and Venegas P~A 2021
  {\em J. Phys.: Cond. Matter\/} {\bf 33} 305801

\bibitem{song_guiding_2020}
Song M, Moon K~W, Yang S, Hwang C and Kim K~J 2020 {\em Appl. Phys. Express\/}
  {\bf 13} 063002

\bibitem{barker_static_2016}
Barker J and Tretiakov O~A 2016 {\em Phys. Rev. Lett.\/} {\bf 116} 147203

\bibitem{legrand_room-temperature_2020}
Legrand W, Maccariello D, Ajejas F, Collin S, Vecchiola A, Bouzehouane K,
  Reyren N, Cros V and Fert A 2020 {\em Nature Mater.\/} {\bf 19} 34--42

\bibitem{duzgun_commensurate_2020}
Duzgun A, Nisoli C, Reichhardt C~J~O and Reichhardt C 2020 {\em Soft Matter\/}
  {\bf 16} 3338--3343

\bibitem{vizarim_guided_2021}
Vizarim N~P, Reichhardt C, Venegas P~A and Reichhardt C~J~O 2021 {\em J. Mag.
  Mag. Mater.\/} {\bf 528} 167710

\bibitem{Velez22}
V{\' e}lez S, Ruiz-G{\' o}mez S, Schaab J, Gradauskaite E, W{\" o}rnle M~S,
  Welter P, Jacot B~J, Degen C~L, Trassin M, Fiebig M and Gambardella P 2022
  {\em Nature Nanotechnol.\/} {\bf 17} 834--841

\bibitem{reichhardt_thermal_2018}
Reichhardt C and Reichhardt C~J~O 2018 {\em J. Phys.: Condens. Matter\/} {\bf
  31} 07LT01

\end{thebibliography}

\end{document}